\documentclass[traditabstract,usenatbib]{aa}  
\usepackage{graphicx}
\usepackage{txfonts}
\usepackage{natbib}
\begin{document}
   \title{The Carina Flare:}

   \subtitle{What can fragments in the wall tell us?\thanks{Based on observations made with ESO telescopes at La Silla Paranal Observatory under programme ID 086.c-0187}}

   \titlerunning{The Carina Flare: What can fragments in the wall tell us?}

   \author{Richard W\"unsch\inst{1}
          \and
          Pavel J\' achym\inst{1}
          \and
          Vojt\v ech Sidorin\inst{1}
          \and 
          So\v na Ehlerov\' a\inst{1}
          \and
          Jan Palou\v s\inst{1}
          \and
          James Dale\inst{2}
          \and
          Joanne R. Dawson\inst{3}
          \and
          Yasuo Fukui\inst{4}
}

   \institute{Astronomical Institute, Academy of Sciences of the Czech Republic, Bocni II 1401--2a, Prague\\
              \email{richard@wunsch.cz}
         \and
             Excellence Cluster `Universe', Boltzmannstr. 2, 85748 Garching,
             Germany
         \and
            School of Mathematics and Physics, University of Tasmania, Sandy Bay
            Campus, Churchill Avenue, Sandy Bay, TAS 7005
         \and
           Department of Physics and Astrophysics, Nagoya University, 
Chikusa-ku, Nagoya, Japan}            

   \date{Received 9 September 2011 / Accepted 18 December 2011}

 
  \abstract
  {$^{13}$CO(J=2--1) and C$^{18}$O(J=2--1) observations of the molecular cloud
  G285.90+4.53 (Cloud~16) in the Carina Flare supershell (GSH287+04-17)
  with the APEX telescope are presented. With an algorithm DENDROFIND we
  identify 51 fragments and compute their sizes and masses. We discuss their
  mass spectrum and interpret it as being the result of the shell fragmentation
  process described by the pressure assisted gravitational instability - PAGI.
  We conclude that the explanation of the clump mass function needs a
  combination of gravity with pressure external to the shell.}


   \maketitle

\label{firstpage}

\section{Introduction} 

Studies of neutral atomic hydrogen (HI) in the Milky Way
\citep{2005A&A...440..775K} and in nearby galaxies \citep{2008AJ....136.2563W}
have discovered shells, fountains and chimneys as ubiquitous fine-scale
features in the HI gas content of galaxies \citep{1979ApJ...229..533H,
2005A&A...437..101E, 2011AJ....141...23B}. In the majority of cases these
structures may be explained by the feedback from young and massive stars.
Stellar radiation, winds and supernovae of  OB associations and of super star
clusters are the main driving force forming the ISM holes and expanding HI
shells. In the HI searches, three kinds of cavity are distinguished: complete
holes surrounded by dense ISM shells, one-sided shells which have blown out of
the galactic disk into the halo in only one direction, and chimneys, in which a
shell has blown out on both sides, puncturing the galactic disk. These latter
blown-out structures provide hot gas and metals to galactic haloes. The chimney
walls formed from the swept-up gas may self-shield against UV photons, form
molecules and cool to very low temperatures. With the help of the combined
action of thermal instabilities \citep{2002ApJ...564L..97K, 2005A&A...433....1A,
2006ApJ...643..245V}, hydrodynamic instabilities \citep{1983ApJ...274..152V,
1991ApJ...368..411R} and gravitational instabilities \citep{1983ApJ...274..152V,
1994ApJ...427..384E, 1994MNRAS.268..291W}, the molecular layers can fragment
into clumps, forming sites of secondary triggered star formation.

In a recent series of papers \citep{2009MNRAS.398.1537D, 2010MNRAS.407.1963W,
2011MNRAS.411.2230D}, we investigated the gravitational fragmentation of a shell
expanding into a high-pressure but low-density medium confining the shell by
thermal pressure but not by ram pressure. We observed that the ambient pressure
was of crucial importance in determining the size scales on which the shell
fragmented. We now seek to confront our theoretical work with observations of
fragments in the walls of the Carina Flare supershell.

The Carina Flare supershell, GSH287+04-17, on the near side of the Carina Arm is
one of the closest examples of an expanding shell reaching high galactic
latitudes in the Milky Way disk. It was discovered in $^{12}$CO(J=1--0) with the
NANTEN telescope by \cite{1999PASJ...51..751F}. \cite{2008MNRAS.387...31D} used
Parkes telescope HI observations to define the Carina Flare as an elliptical
shell approximately $230\times 360$~pc centred $\sim 250$~pc above the Galactic
Plane. They computed its HI mass as $7\pm 3\times 10^{5}$~M$_{\odot}$ and its
molecular hydrogen mass as $2.0 \pm 0.6 \times 10^{5}$~M$_{\odot}$, and
estimated the shell's kinetic energy as $10^{51}$~erg.

\cite{2008PASJ...60.1297D} used the NANTEN millimetre telescope to study the
molecular content in more detail.  In $^{12}$CO(J=1--0) and $^{13}$CO(J=1--0)
observations, they identified $\sim$220 clouds down to a completeness limit of
$\sim60$~M$_{\odot}$. They derived fragment mass functions with slopes of $\sim
-1.5$ and found that all of their clumps had strongly subvirial masses. This
latter result immediately raises the question of what instability
caused the Carina Flare to fragment, since the action of the gravitational
thin-shell instability would be expected to produce fragments with virial or
super-virial masses.

In a parsec-resolution study of the HI and CO distributions in GSH287+04-17,
\citet{2011ApJ...728..127D} distinguished between different origins of the
molecular clouds in the Carina Flare walls. They identified the remnants of
pre-existing clouds that have been engulfed by the hot medium in the expanding
HI hole, as well as locations where molecular clouds may have condensed from the
swept-up atomic medium collected into the shell walls. Of the latter, one of the
most compelling examples is the molecular cloud G285.90+4.53 (hereafter
Cloud~16, as in \citealt{2008PASJ...60.1297D}), which lies approximately halfway
up the western side of the shell, $\sim 250$~pc above the Galactic Plane in a
region named ``the approaching limb complex.''

The $^{12}$CO(J=1--0), $^{13}$CO(J=1--0) and C$^{18}$O(J=1--0) lines in this
region have also been observed by the Mopra 22m telescope, providing detailed
information on the distribution and properties of the molecular gas to a
resolution of $0.6$~pc, and confirming massive star formation activity in the
southernmost regions of the cloud \citep{2011arXiv1108.3882D}. In this
paper, we report the first results of our new high-resolution
observations of $^{13}$CO(J=2--1) and C$^{18}$O(J=2--1) in the Carina Flare
Cloud~16 with the APEX telescope\footnote{This publication is based on data
acquired with the Atacama Pathfinder Experiment (APEX). APEX is a collaboration
between the Max-Planck-Institute fur Radioastronomie, the European Southern
Observatory, and Onsala Space Observatory.}. With our signal-to-noise ratio,
which is much better compared to the MOPRA data, and a spatial resolution of
$\sim 0.34$ pc, we identify 51 clumps. We discuss their sizes and masses, some
of which are close to virial. We also discuss the possible origin of their mass
distribution, which can not be created by a pure gravitational instability.

\section{Observations of the Carina Flare Cloud~16}

\begin{figure*}
\centering
\includegraphics[width=\textwidth]{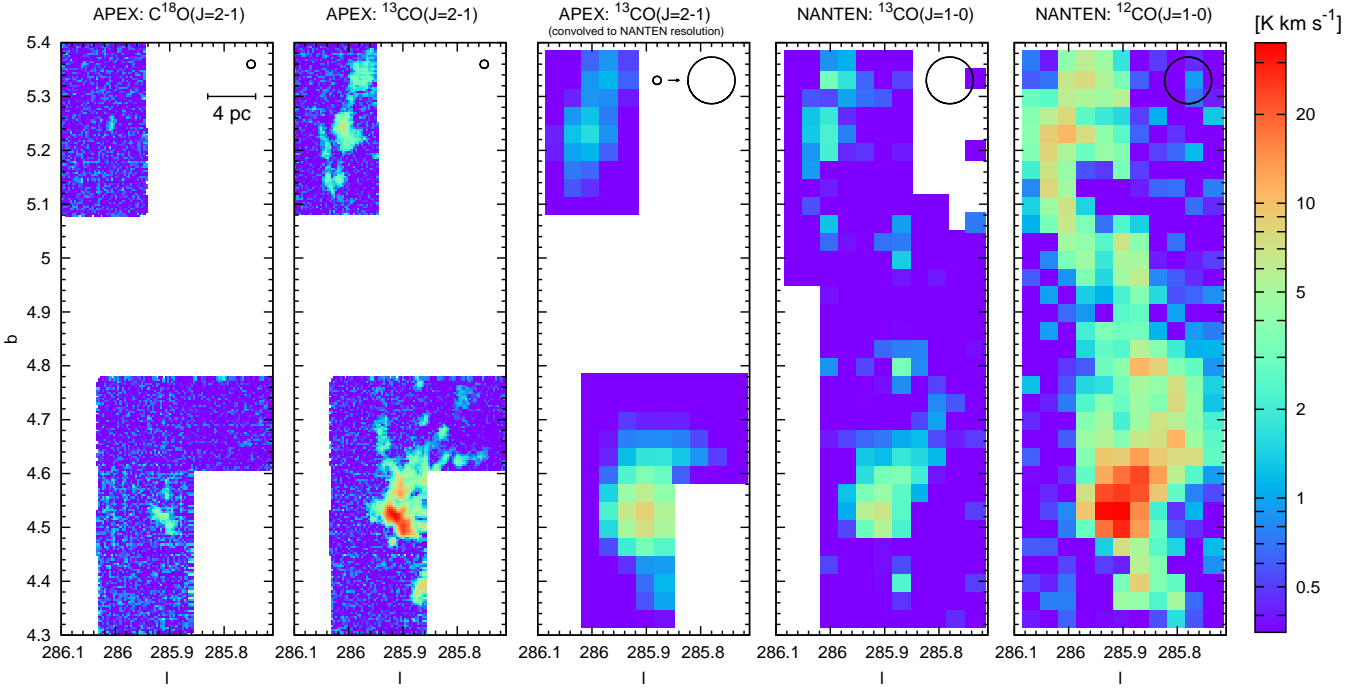}
\caption{Velocity integrated C$^{18}$O(J=2--1) and $^{13}$CO(J=2--1) emission
observed with APEX (left two panels) versus  $^{13}$CO(J=1--0) and
$^{12}$CO(J=1--0) emission observed with NANTEN (right two panels).
The middle panel shows the APEX $^{13}$CO(J=2--1) convolved to the
resolution of NANTEN. The circle in the upper right corner
corresponds to the beam FWHP (full width at half power). 
} 
\label{apexnanten}
\end{figure*}

The observations were carried out with the Atacama Pathfinder
EXperiment (APEX) 12-m diameter antenna in December 2010 and January 2011. Using
the APEX-1 receiver of the Swedish Heterodyne Facility Instrument
(SHeFI), the observations were done simultaneously at the frequencies of
the $^{13}$CO(J=2--1) ($\nu_{\rm rest} = 220.399$~GHz) and
C$^{18}$O(J=2--1) ($\nu_{\rm rest} = 219.560$~GHz) lines. The Fast Fourier
Transform Spectrometer backend was used with a bandwidth of 1~GHz
divided into 2048 channels, corresponding to a velocity resolution of
0.66~km\,s$^{-1}$. At 220~GHz, the half-power beam width of the telescope is
about $27"$ which at the adopted distance of the Carina Flare of
2.6~kpc corresponds to about 0.34~pc. The system temperatures were
typically in the range 150--170~K; only a very small part of the
data had significantly poorer quality with $T_\mathrm{sys}$ up to 650~K.

To map the brightest parts of Cloud~16, we observed six $10'\times 10'$
rectangular areas -- four located in an ``L''-shaped lower region, and two in an
upper region, separated by about $18'$ (see Figure~\ref{apexnanten}). We used
On-The-Fly (OTF) mode which allows the telescope to scan continuously along one
direction while reading out data every $10"$ after 1~second integration
intervals. The scanning pattern used was a ``zigzag'' through the regions
which were tilted by $26^\circ$ with respect to the RA axis. The second row was
scanned in the opposite direction, and so on, with $10"$ spacing. This
pixel size ($10"\times 10"$) corresponds to the spatial resolution
$0.18\,\mathrm{pc}\, \times\, 0.18\,\mathrm{pc}$ at the distance of Carina Flare
($2.6$~kpc). Each map (except the top box of the upper region) was observed
twice in two perpendicular directions. The neighbouring regions overlapped by
one beam width. The duration of one map with OTF overheads was about 2.5 hours.
The total observing time was then 22.2~hrs of which about 2~hrs were lost.
The observing conditions were good to excellent with pwv less than 0.6~mm.

The data were reduced according to the standard procedure using CLASS from the
GILDAS software package developed at IRAM. Bad scans were flagged and emission
line-free channels were used to subtract in most cases first-order baselines.
Some spectra showed a wide residual double sinusoidal variation caused probably
by a vibration of either the cover of the Cassegrain cabin entrance window or
the cold head of the closed-cycle cooling machine \citep{2006A&A...454L.111D}.
We used the original velocity resolution of $\sim 0.66$~km\,s$^{-1}$. The
corrected antenna temperatures, $T^*_{\rm A}$, provided by the APEX calibration
pipeline \citep{2010SPIE.7737E..37D}, were converted to main-beam brightness
temperature by $T_{\rm b}=T^*_{\rm A} / \eta_{\rm mb}$, using the main beam
efficiency $\eta_{\rm mb} =0.75$ at $220$~GHz. Finally, FITS cubes were
produced by CLASS for individual observed regions and stitched together manually
using a Python routine. RMS noise levels of better than $0.2$~K per
$0.66$~km\,s$^{-1}$ channels were obtained. 


The high angular resolution of APEX allows us in principle to observe
objects of mass $\sim 0.6$~M$_{\odot}$, well into the regime of stellar masses.
In Figure \ref{apexnanten}, we compare APEX and NANTEN velocity channel maps to
illustrate the improvement in resolution afforded by APEX. Our observations show
clearly that Cloud~16 is composed of irregularly shaped clumps connected by
filaments, and that structure exists on scales considerably smaller than were
accessible with NANTEN.

\section{Identification of Clumps}
\label{sec:dendrofind}

The structure of the molecular gas in galaxies is hierarchical.
\citet{2008ApJ...679.1338R} showed that it is useful to display its essential
features as dendrograms -- graphical representations of the topology of the
isosurfaces as a function of contour level. We have developed a simple algorithm
called DENDROFIND which creates dendrograms as a natural by-product of the
clump-finding process. The algorithm is similar to the well-known CLUMPFIND
\citep{1994ApJ...428..693W} but its results are less dependent on
technical parameters (for instance, the brightness temperature difference between
contours). Another advantage comparing to CLUMPFIND is that DENDROFIND
always produces contiguous clumps (CLUMPFIND may lead to discontiguous structures
identified as a single clump in extreme cases).

The DENDROFIND algorithm operates on a position-position-velocity datacube of
the brightness temperature $T_\mathrm{b}$. The main loop gradually decreases the
temperature level \texttt{Tl} starting from the maximum brightness temperature
in the datacube, \texttt{Tmax}, going down to \texttt{Tcutoff} set by the user.
In each iteration, pixels with $T_\mathrm{b} >$~\texttt{Tl} are assigned to existing
clumps, or new clumps are created and the pixels are assigned to them.
Furthermore, information about connections among clumps on level \texttt{Tl} is
evaluated and recorded. This information is necessary (and sufficient) to plot
the dendrogram at the end of the algorithm. Each clump created by the algorithm
has to fulfil the following three conditions: (i) it is a contiguous area in the
PPV space; (ii) it consists of at least \texttt{Npxmin} pixels (\texttt{Npxmin}
is another user defined parameter); and (iii) the difference between the clump
peak temperature and the temperature at which the clump connects to other clumps
has to be at least \texttt{dTleaf}. This parameter effectively controls whether
a clump should be split into several objects, or just regarded as a single
object with unresolved substructure.

In summary, the algorithm is controlled by four parameters: \texttt{Nlevels}
(number of temperature levels), \texttt{Npxmin} (minimum number of pixels of
a clump), \texttt{dTleaf} (minimum difference between the clump peak
temperature and the temperature at which the clump connects to other clumps) and
\texttt{Tcutoff} (minimum $T_\mathrm{b}$ considered for assignment to clumps). It is
always possible to find \texttt{Nlevels} such that using higher values would
lead to identical results. The DENDROFIND algorithm is described in detail
in Appendix \ref{ap:df}.

\begin{figure}[t]
\centering
\includegraphics[width=\columnwidth]{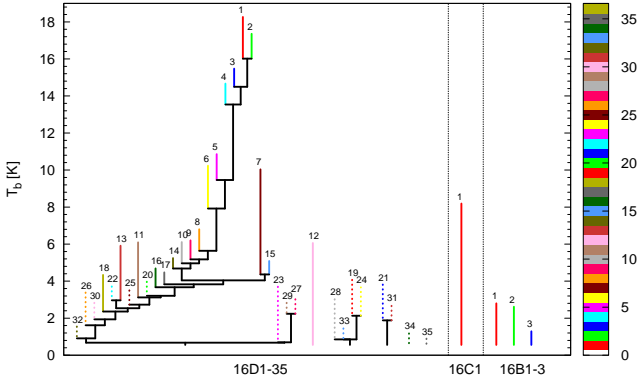}
\caption{Dendrogram of the clumps in the $^{13}$CO(J=2--1) emission in the lower
region. Colors of lines representing clumps are the same as colors of regions
showing clumps in Figures \ref{clumps_ex} and \ref{clumps1}--\ref{clumps3}.
Solid lines and filled regions (in Figures \ref{clumps_ex} and
\ref{clumps1}--\ref{clumps3}) show clumps 16D1--18, 16C1 and 16B1--3, dashed
lines and dotted regions (in Figures \ref{clumps_ex} and
\ref{clumps1}--\ref{clumps3}) show clumps 16D19--35 (see beginning of
\S\ref{sec:clump_properties} for the nomenclature definition).}
\label{dendro1}
\end{figure}

\begin{figure}[t]
\centering
\includegraphics[width=\columnwidth]{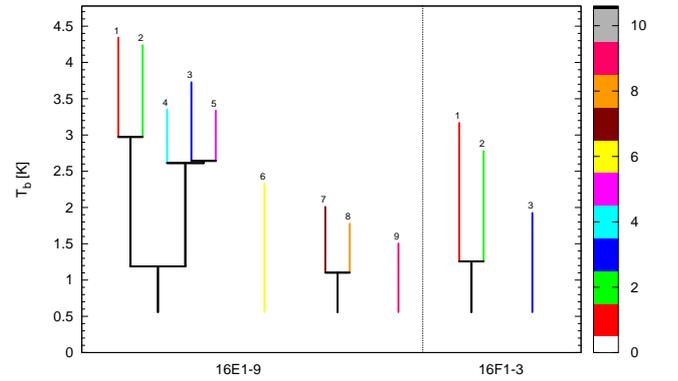}
\caption{Dendrogram of the clumps in the $^{13}$CO(J=2--1) emission in the upper
region. Colors of lines representing clumps are the same as colors of regions
showing clumps in Figures \ref{clumps4}--\ref{clumps5}.}
\label{dendro2}
\end{figure}

\begin{figure*}
\centering
\includegraphics[width=\textwidth]{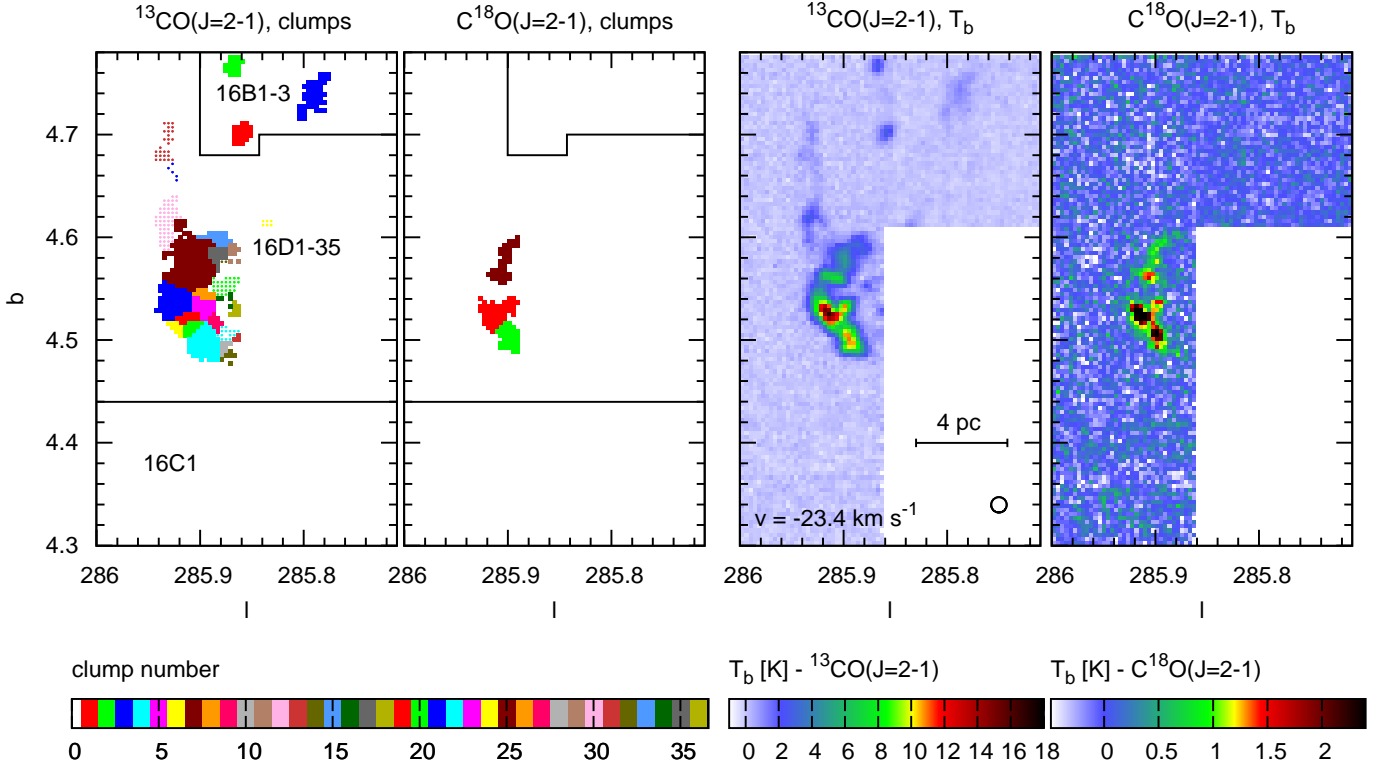}
\caption{Clumps identified in the lower region (left two panels) and
corresponding brightness temperature (right two panels) in the velocity channel
$v = 23.4$~km\,s$^{-1}$. Filled regions show clumps 1-18, dotted regions show clumps
19-35. Colors of clumps correspond to colors of dendrogram leaves in
Figure~\ref{dendro1}.}
\label{clumps_ex}
\end{figure*}

In this work, DENDROFIND has been applied to the $^{13}$CO(J=2--1) and
C$^{18}$O(J=2--1) data of the observed part of the Carina Flare Cloud~16. The
parameters were \texttt{Nlevels} = 1000, \texttt{Npxmin} = 5 and \texttt{dTleaf}
= \texttt{Tcutoff} = $3\sigma_\mathrm{noise}$ where $\sigma_\mathrm{noise}$ is
the standard deviation of the noise in the data cube. The resulting dendrograms
are seen in Figures~\ref{dendro1} and \ref{dendro2}. In the $^{13}$CO data, we
identified 51 clumps, which are listed in Tables \ref{table:13CO_clumps_lower}
and \ref{table:13CO_clumps_upper}. Some of these clumps were also found in the
C$^{18}$O data (see Tables~\ref{table:C18O_clumps_lower} and
\ref{table:C18O_clumps_upper}). Figure~\ref{clumps_ex} shows an example radial
velocity channel map with the identified clumps. All radial velocity channel
maps with clumps are shown in the online appendix~\ref{ap:figs}.

%
%

\section{Clump Sizes, Velocity Dispersions and Masses}
\label{sec:clump_properties}


Tables~\ref{table:13CO_clumps_lower}--\ref{table:C18O_clumps_upper} list
properties of clumps identified in $^{13}$CO(J=2--1) and C$^{18}$O(J=2--1) APEX
observations. ``Label'' is the clump-identifier based on the list given by
\citet{2008PASJ...60.1297D}. The first number and letter is identical to the
list of \citet{2008PASJ...60.1297D}, e.g. ``16D'', then a number denoting the
substructure which we resolve is appended.
%
``FWHM$_x$'' is the full width at half maximum of the clump in the direction
of coordinate $x$, where $x \in \{ l,\, b,\, v \equiv v_{\mathrm{LSR}} \}$.
FWHM$_x$ is given by the formula

\begin{equation}
  \label{eq:FWHMx}
  \mathrm{FWHM}_x = 2 \sqrt{ 2 \ln 2 } \; \sqrt{\sigma_x^2 + \left(\frac{\Delta x}{2 \sqrt{ 2 \ln 2 }}\right)^2}
\end{equation}
where the second term under the square root represents the pixel resolution term
with $\Delta x$ being the pixel width in the $x$ coordinate.
This term ensures the FWHM is not smaller than the pixel size. 
If the pixel size goes to zero the FWHM approaches a Gaussian FWHM.
$\sigma_x^2$ is the weighted variance, with the brightness
temperature $T_\mathrm{b}$ as the weight:

\begin{equation}
  \label{eq:sigma2}
  \sigma_x^2 = \overline{x^2} - \overline{x}^2
             = \sum_{\mathrm{px}}^{} \frac{x^2 T_\mathrm{b}}{T_{\mathrm{tot}}}
               - \left( \sum_{\mathrm{px}}^{} \frac{xT_\mathrm{b}}{T_{\mathrm{tot}}} \right)^2
\end{equation}

\begin{equation}
  \label{eq:Ttot}
  \mathrm{where} \quad
  T_{\mathrm{tot}} = \sum_{\mathrm{px}} T_\mathrm{b}
\end{equation}
The summation in the equations above is over all a given clump's pixels.
%
$R$ is the clump equivalent radius defined as

\begin{equation}
  \label{eq:R}
  R = \sqrt{A_{\mathrm{cl}}/\pi}
\end{equation}
where $A_{\mathrm{cl}}$ is the clump projected area in the sky ($lb$-plane).
%
$T_{\mathrm{peak}}$ is the brightness temperature of the brightest pixel of the
clump, $l_\mathrm{peak}$, $b_\mathrm{peak}$ are its J2000 galactic
coordinates and $v_\mathrm{peak}$ is its LSR radial velocity.
%
The next column, $m$, is the total clump mass computed as

\begin{equation}
  \label{eq:M}
  m = \frac{m_{\mathrm{H_2}}}{X},
  \qquad \mathrm{where} \quad
  X = 0.735
\end{equation}
and $m_{\mathrm{H_2}}$ is the total clump mass in the form of molecular hydrogen
computed from the formula

\begin{equation}
  \label{eq:MH2}
  m_{\mathrm{H_2}} = \mu_{\mathrm{H_2}} \sum_{\mathrm{px}} N_{\mathrm{H_2}} \cdot A_{\mathrm{px}}
\end{equation}
%
where $\mu_{\mathrm{H_2}}$ is the mass of one hydrogen molecule,
$N_{\mathrm{H_2}}$ is the column density of molecular hydrogen in a given pixel
and $A_{\mathrm{px}}$ is the pixel area. The column density $N_{\mathrm{H_2}}$
is derived from the column densities of the $^{13}$CO and C$^{18}$O 
molecules assuming the standard abundances \citep{1998gaas.book.....B}:
$N_\mathrm{H_2} = 6.5 \times 10^{5} N_\mathrm{^{13}CO}$,
$N_\mathrm{H_2} = 5   \times 10^{6} N_\mathrm{C^{18}O}$.
These values are in agreement with recent measurements and models; see, e.g.,
\citet{2007ApJS..168...58S, 2009A&A...503..323V, 1999RPPh...62..143W}
and the citations therein. Abundances, however, vary for different
regions/clouds and depend also on the physical conditions inside individual
clumps.  Although this variation is a very fundamental problem,
a comprehensive discussion is beyond the scope of this paper.

The column densities of both $^{13}$CO and C$^{18}$O are, in the case of
$J=2-1$ lines, computed from the same formula \citep[14.54]{1996tra..book.....R}

\begin{equation}
\label{eq:NCO}
\frac{N}{\mathrm{cm}^{-2}} =
1.51 \times 10^{14} \cdot
\frac{\mathrm{e}^{5.3/T_{\mathrm{ex}}} \; T_{\mathrm{ex}} }{1-\mathrm{e}^{-10.6/T_{\mathrm{ex}}}} \;
\int
\tau \;
\mathrm{d} v
\end{equation}
This formula assumes local thermodynamic equilibrium characterised by the
excitation temperature $T_{\mathrm{ex}}$, which we assume to be $25$~K. This
value is slightly higher than that usually assumed in such analyses (between
$10$\,K and $20$\,K) and the value we use in \S\ref{sec:pagi} to analyze
the fragmentation process ($15$\,K). The reason for this assumption is
that the brightest pixel in our data is over $18\;\mathrm{K}$ (see
$T_\mathrm{peak}$ of clump 16D1), which puts the lower limit to the excitation
temperature somewhere above $20\;\mathrm{K}$. This is so because the excitation
temperature has to be higher than the brightness temperature, assuming local
thermodynamic equilibrium. Note that the column density, and thus the mass,
is only higher by about $5\;\%$ for most clumps for an excitation temperature
of $25\;\mathrm{K}$ compared to $20\;\mathrm{K}$. In our case of discrete
pixels, $\mathrm{d}v$ corresponds to $\Delta v$, where $\Delta v$ is the pixel
width in coordinate $v$ (radial velocity channel width, $\Delta v \approx 0.66
\; \mathrm{km\,s^{-1}}$). $\tau$ is the optical depth of $^{13}$CO(J=2--1) or
C$^{18}$O(J=2--1) given by the formula

\begin{equation}
  \label{eq:tauCOpix}
  \tau = -\ln \left[ 1 - \frac{T}{10.6} \cdot 
  \left( \frac{1}{\exp(10.6/T_{\mathrm{ex}}) - 1} - 0.02 \right)^{-1} \right]
\end{equation}
which we derived from formula 14.46 of \citet{1996tra..book.....R}.

In addition to uncertainties caused by departures from LTE, the accuracy of the
clump mass $m$ depends on the data noise $\sigma_T$. In our datasets,
$\sigma_T$ is between $0.18\;\mathrm{K}$ and $0.20\;\mathrm{K}$, which
corresponds to an error in the determination of the pixel mass $\Delta
m_\mathrm{px} = 0.04$~M$_\odot$ for the $^{13}$CO(J=2--1) line and $\Delta
m_\mathrm{px} = 0.3$~M$_\odot$ for the C$^{18}$O(J=2--1) line. Assuming the
noise is Gaussian and neglecting the correlation of pixels within the
beam, the error in determination of the clump mass (given in 11th column of
Tables~\ref{table:13CO_clumps_lower}--\ref{table:C18O_clumps_upper}) is
\begin{equation}
\label{deltam}
\Delta m = \sqrt{N_\mathrm{px}} \Delta m_\mathrm{px}
\end{equation}
where $N_\mathrm{px}$ is the number of pixels in the clump.
Since each clump has to have at least 5 pixels, the minimum clump mass is
$0.6$~M$_\odot$ and $4.5$~M$_\odot$ for $^{13}$CO and C$^{18}$O
data, respectively.

To estimate the overall error in the mass determination, we compare our APEX
$^{13}$CO(2--1) observation with the NANTEN $^{13}$CO(1--0) observation by
convolving the APEX data to the resolution of NANTEN (see the middle
panel of Figure~\ref{apexnanten}). Both distributions of the brightness
temperature show a very good agreement. The total mass detected in both lower
and upper regions in the convolved APEX data is $\sim 2.8\times 10^3$~M$_\odot$,
which differs by only $15$~\% from the total mass detected in the same two
regions in the NANTEN data ($\sim 2.4\times 10^3$~M$_\odot$). These values also
correspond well to the total mass in the identified $^{13}$CO clumps (see
Tables~\ref{table:13CO_clumps_lower} and \ref{table:13CO_clumps_upper}) which
is $\sim 2.4\times 10^3$~M$_\odot$.

As another test of reliability of the clump mass derivation, we compare masses
obtained from the two lines observed by APEX. We select areas identified as
clumps in the C$^{18}$O data (Tables~\ref{table:C18O_clumps_lower} and
\ref{table:C18O_clumps_upper}) and calculate masses contained in them from both
C$^{18}$O and $^{13}$CO datasets. The results of the comparison are shown in
Table~\ref{tab_entangled_masses}. There are four C$^{18}$O clumps in the lower
region and two in the upper region. In the lower region, the mean difference
between C$^{18}$O and $^{13}$CO masses is about $12$~\%, while in the upper
region, the differences for the two clumps are about $70$~\%. Part of these
differences can be due to varying isotope abundances, different excitation
temperatures in the lower and upper regions, and departures from LTE. From this
comparison and from the comparison with the NANTEN data, we conclude that
systematic errors in our determination of clump masses are probably less
than a factor of two.

The 12th column in
Tables~\ref{table:13CO_clumps_lower}--\ref{table:C18O_clumps_upper},
$m_{\mathrm{vir}}$, is the estimated clump virial mass, given by the formula
\citep{1988ApJ...333..821M}

\begin{equation}
  \label{eq:Mvir}
  m_{\mathrm{vir}} = 190 \cdot R \cdot {\mathrm{FWHM}_v}^2
\end{equation}

\noindent The formula above assumes spherical clumps with a density profile $\propto 1/r$.
%
The virial mass error $\Delta m_{vir}$ is computed from the first order Taylor expansion
assuming the clump's area and FWHM$_v$ errors of one pixel.
The last column in Tables
\ref{table:13CO_clumps_lower}--\ref{table:C18O_clumps_upper} gives the ratio of
the total clump mass derived from the brightness temperature and the clump
virial mass.


\begin{table*}
  \centering
\begin{tabular}{lrrrrrrrrrrrrr}
  Label &
  $\frac{\mathrm{FWHM}_{x}}{\mathrm{pc}}$ &
  $\frac{\mathrm{FWHM}_{y}}{\mathrm{pc}}$ &
  $\frac{\mathrm{FWHM}_{v}}{\mathrm{km\;s^{-1}}}$ &
  $\frac{R}{\mathrm{pc}}$ &
  $\frac{T_{\mathrm{peak}}}{\mathrm{K}}$ &
  $\frac{l_\mathrm{peak}}{\mathrm{deg}}$ &
  $\frac{b_\mathrm{peak}}{\mathrm{deg}}$ &
  $\frac{v_\mathrm{peak}}{\mathrm{km\;s^{-1}}}$ &
  $\frac{m}{\mathrm{M}_\odot}$ &
  $\frac{\Delta m}{\mathrm{M}_\odot}$ &
  $\frac{m_{\mathrm{vir}}}{\mathrm{M}_\odot}$ &
  $\frac{\Delta m_{\mathrm{vir}}}{\mathrm{M}_\odot}$ &
  $\frac{m}{m_{\mathrm{vir}}}$
  \\
  \hline\hline
  16D1  &   0.59 &   0.37 &   1.98 &   0.4 &   18.3 & 285:54:51.7 & 4:31:16.6 & -22.75 &  144.6 & 0.33 &  330.7 &    221.4 &    0.437 \\
  16D2  &   0.51 &   0.51 &   1.91 &   0.5 &   17.4 & 285:54:37.4 & 4:31:02.3 & -22.75 &  119.6 & 0.37 &  315.2 &    218.9 &    0.379 \\
  16D3  &   1.08 &   0.90 &   2.07 &   1.1 &   15.5 & 285:55:05.9 & 4:31:45.2 & -23.41 &  312.2 & 0.79 &  895.8 &    574.0 &    0.348 \\
  \dots &        &        &        &       &        &     \dots   &           &        &        &      &        &          &  \dots   \\
\end{tabular}
  \caption{$^{13}$CO(J=2--1) clump properties -- lower region. The meaning
  of the columns and the means by which the values were calculated are explained in
  \S\ref{sec:clump_properties}. Clumps marked by the asterisk at their label are
  excluded from the CMF, because they lie at the border of the observed area.
  Full version of this table is available in the online appendix -- see
  Table~\ref{table:13CO_clumps_lower_full}.
  }
  \label{table:13CO_clumps_lower}
\end{table*}


\begin{table*}
  \centering
\begin{tabular}{lrrrrrrrrrrrrr}
  Label &
  $\frac{\mathrm{FWHM}_{x}}{\mathrm{pc}}$ &
  $\frac{\mathrm{FWHM}_{y}}{\mathrm{pc}}$ &
  $\frac{\mathrm{FWHM}_{v}}{\mathrm{km\;s^{-1}}}$ &
  $\frac{R}{\mathrm{pc}}$ &
  $\frac{T_{\mathrm{peak}}}{\mathrm{K}}$ &
  $\frac{l_\mathrm{peak}}{\mathrm{deg}}$ &
  $\frac{b_\mathrm{peak}}{\mathrm{deg}}$ &
  $\frac{v_\mathrm{peak}}{\mathrm{km\;s^{-1}}}$ &
  $\frac{m}{\mathrm{M}_\odot}$ &
  $\frac{\Delta m}{\mathrm{M}_\odot}$ &
  $\frac{m_{\mathrm{vir}}}{\mathrm{M}_\odot}$ &
  $\frac{\Delta m_{\mathrm{vir}}}{\mathrm{M}_\odot}$ &
  $\frac{m}{m_{\mathrm{vir}}}$
  \\
  \hline\hline
  16D1 &   1.04 &   0.79 &   1.73 &   0.8 &    3.2 & 285:54:51.7 & 4:31:16.6 & -22.75 &  296.7 & 3.55 & 481.9 &    368.9 &    0.616 \\
  16D2 &   0.59 &   0.80 &   1.51 &   0.6 &    2.5 & 285:53:54.6 & 4:30:19.3 & -23.41 &  168.0 & 2.77 & 277.8 &    244.4 &    0.605 \\
  16D7 &   0.64 &   1.29 &   1.01 &   0.7 &    2.5 & 285:54:08.4 & 4:34:07.7 & -22.75 &  139.1 & 2.72 & 138.8 &    181.8 &    1.002 \\
  16C1 &   0.41 &   0.58 &   0.66 &   0.4 &    1.4 & 285:51:47.0 & 4:23:52.9 & -25.40 &   22.6 & 1.12 &  31.8 &     63.6 &    0.711 \\
\end{tabular}
  \caption{C$^{18}$O(J=2--1) clump properties -- lower region. The meaning
  of the columns and the means by which the values were calculated are explained in
  \S\ref{sec:clump_properties}.}
  \label{table:C18O_clumps_lower}
\end{table*}


\begin{table*}
  \centering
\begin{tabular}{lrrrrrrrrrrrrr}
  Label &
  $\frac{\mathrm{FWHM}_{x}}{\mathrm{pc}}$ &
  $\frac{\mathrm{FWHM}_{y}}{\mathrm{pc}}$ &
  $\frac{\mathrm{FWHM}_{v}}{\mathrm{km\;s^{-1}}}$ &
  $\frac{R}{\mathrm{pc}}$ &
  $\frac{T_{\mathrm{peak}}}{\mathrm{K}}$ &
  $\frac{l_\mathrm{peak}}{\mathrm{deg}}$ &
  $\frac{b_\mathrm{peak}}{\mathrm{deg}}$ &
  $\frac{v_\mathrm{peak}}{\mathrm{km\;s^{-1}}}$ &
  $\frac{m}{\mathrm{M}_\odot}$ &
  $\frac{\Delta m}{\mathrm{M}_\odot}$ &
  $\frac{m_{\mathrm{vir}}}{\mathrm{M}_\odot}$ &
  $\frac{\Delta m_{\mathrm{vir}}}{\mathrm{M}_\odot}$ &
  $\frac{m}{m_{\mathrm{vir}}}$
  \\
  \hline\hline
  16E1 &   0.94 &   1.66 &   1.20 &   1.2 &    4.3 & 286:00:20.5 & 5:14:31.2 & -24.74 &  121.0 & 0.69 &  335.8 &    372.1 &    0.360 \\
  16E2 &   1.32 &   0.99 &   0.88 &   1.0 &    4.2 & 285:59:51.5 & 5:13:20.1 & -24.08 &   57.5 & 0.47 &  153.3 &    231.1 &    0.375 \\
  16E3 &   1.32 &   1.29 &   1.13 &   1.1 &    3.7 & 285:59:10.6 & 5:19:45.3 & -24.08 &   56.4 & 0.54 &  270.0 &    317.8 &    0.209 \\
 \dots &        &        &        &       &        &     \dots   &           &        &        &      &        &          &  \dots   \\
\end{tabular}
  \caption{$^{13}$CO(J=2--1) clump properties -- upper region. The meaning
  of the columns and the means by which the values were calculated are explained in
  \S\ref{sec:clump_properties}. Clumps marked by the asterisk at their label are
  excluded from the CMF, because they lie at the border of the observed area.
  Full version of this table is available in the online appendix -- see
  Table~\ref{table:13CO_clumps_upper_full}.
  }
  \label{table:13CO_clumps_upper}
\end{table*}


\begin{table*}
  \centering
\begin{tabular}{lrrrrrrrrrrrrr}
  Label &
  $\frac{\mathrm{FWHM}_{x}}{\mathrm{pc}}$ &
  $\frac{\mathrm{FWHM}_{y}}{\mathrm{pc}}$ &
  $\frac{\mathrm{FWHM}_{v}}{\mathrm{km\;s^{-1}}}$ &
  $\frac{R}{\mathrm{pc}}$ &
  $\frac{T_{\mathrm{peak}}}{\mathrm{K}}$ &
  $\frac{l_\mathrm{peak}}{\mathrm{deg}}$ &
  $\frac{b_\mathrm{peak}}{\mathrm{deg}}$ &
  $\frac{v_\mathrm{peak}}{\mathrm{km\;s^{-1}}}$ &
  $\frac{m}{\mathrm{M}_\odot}$ &
  $\frac{\Delta m}{\mathrm{M}_\odot}$ &
  $\frac{m_{\mathrm{vir}}}{\mathrm{M}_\odot}$ &
  $\frac{\Delta m_{\mathrm{vir}}}{\mathrm{M}_\odot}$ &
  $\frac{m}{m_{\mathrm{vir}}}$
  \\
  \hline\hline
   16E1 &   0.53 &   0.67 &   0.83 &   0.5 &    1.5 & 286:00:20.6 & 5:14:45.5 & -24.74 &   36.9 & 1.53 &  63.9 &    102.1 &    0.577 \\
   16E2 &   0.46 &   0.37 &   0.82 &   0.3 &    1.6 & 286:00:20.5 & 5:14:17.0 & -24.74 &   19.8 & 1.04 &  41.4 &     66.7 &    0.479 \\
\end{tabular}
  \caption{C$^{18}$O(J=2--1) clump properties -- upper region. The meaning
  of the columns and the means by which the values were calculated are explained in
  \S\ref{sec:clump_properties}. }
  \label{table:C18O_clumps_upper}
\end{table*}

\begin{table}
  \centering
  \begin{tabular}{lrrrrr}
    Label &
    $\frac{T_\mathrm{peak}^{(13)}}{\mathrm{K}}$ &
    $\frac{T_\mathrm{peak}^{(18)}}{\mathrm{K}}$ &
    $\frac{m^{(13)}}{\mathrm{M}_\odot}$ &
    $\frac{m^{(18)}}{\mathrm{M}_\odot}$ &
    $\frac{m^{(13)}-m^{(18)}}{m_\mathrm{mean}}$ \\
    \hline\hline
    \\
    \multicolumn{6}{l}{Lower region} \\
    \hline
    16D1  & 18.3 &  3.2 & 367.2 & 296.7 & $+0.21$ \\
    16D2  & 14.7 &  2.5 & 194.5 & 168.0 & $+0.14$ \\
    16D7  & 10.0 &  2.5 & 125.6 & 139.1 & $-0.10$ \\
    16C1  &  8.2 &  1.4 &  23.4 &  22.6 & $+0.03$ \\
    \\
    \multicolumn{6}{l}{Upper region} \\
    \hline
    16E1  &  4.0 &  1.5 &  16.5 &  36.9 & $-0.76$ \\
    16E2  &  4.3 &  1.6 &  10.0 &  19.8 & $-0.66$ \\
  \end{tabular}
  \caption{Comparison of C$^{18}$O clump masses derived from C$^{18}$O (`18')
  and $^{13}$CO (`13') brightness temperature maps.}
  \label{tab_entangled_masses}
\end{table}

\section{Discussion}
\label{sec:disc}
As seen in Tables~\ref{table:13CO_clumps_lower}--\ref{table:C18O_clumps_upper},
the observed masses of individual clumps are, compared to
\cite{2008PASJ...60.1297D}, significantly closer to virial, but they are still
lower than the virial estimates. This difference could mean that besides
the Jeans gravitational instability, there are other forces helping gravity
during the fragmentation process. Another explanation may be that we are
still limited in angular and velocity resolution, so that some of our clumps may
be resolvable still further into smaller denser supervirial objects. 
In particular, the relatively coarse velocity resolution is the main source of
uncertainty in determining virial masses. This may be checked with future even
higher resolution observations with ALMA interferometer.

\subsection{Clump mass function - CMF}
\label{CMF}

\begin{figure}
\centering
\includegraphics[width=\columnwidth]{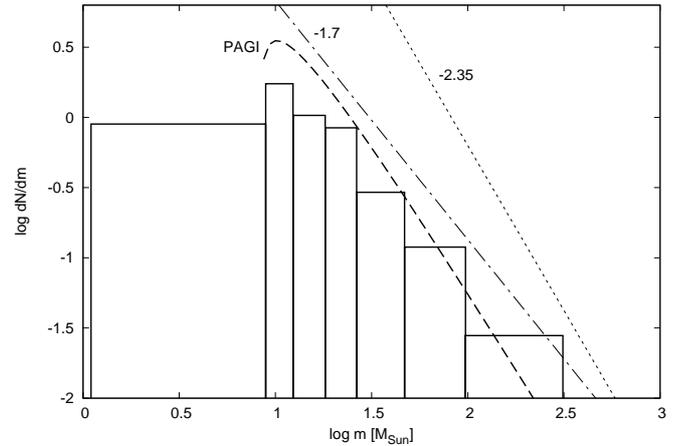}
\caption{Mass spectrum of fragments identified in our $^{13}$CO(J=2--1)
observations (solid line). Following the prescription of
\citet{2005ApJ...629..873M} to minimize errors in histograms with a small number
of objects, we set bin widths so that each bin includes approximately the same
number of objects (six or seven). The dashed line shows the mass spectrum
calculated from the PAGI dispersion relation (see equation \ref{eq:PAGI:ms}) for
$\Sigma = 9\times 10^{-4}$~g\,cm$^{-2}$, $T = 15$~K and $P_{_{\rm EXT}} =
1.4\times 10^{-12}$~dyne\,cm$^{-2}$ (parameters determined in \S\ref{sec:pagi}).
Dash-dotted and dotted lines show power law with slopes $-1.7$ and $-2.35$,
respectively.
}
\label{fig:ms}
\end{figure}

We have determined the mass function of clumps (CMF) identified in
$^{13}$CO(J=2--1) observations of both regions. Clumps lying at the border of
the observed area having $T_\mathrm{peak}-T_\mathrm{border} >$ \texttt{dTleaf},
where $T_\mathrm{border}$ is the maximum brightness temperature of the clump at
the border of the observed area, are excluded from the CMF, because their parts
lying outside of the observed region may substantially contribute to their
masses. There are 7 such clumps and they are marked by asterisks in
Tables~\ref{table:13CO_clumps_lower}--\ref{table:C18O_clumps_upper}. The CMF of
the remaining 44 clumps is shown by Figure~\ref{fig:ms}. 
The maximum of the distribution is around $m_\mathrm{CMF}\sim 10$~M$_\odot$. The
slope of the CMF is compared to the power law with the ``Salpeter'' slope
$-2.35$ (dotted line) and a power law with slope $-1.7$ (dash-dotted line)
found by \citet{1998A&A...329..249K} for clumps in several molecular clouds. The
dashed line shows the mass spectrum obtained from the Pressure Assisted
Gravitational Instability (PAGI; see \S\ref{sec:pagi}). 

The slope of the CMF agrees well with the $-1.7$ power law. The agreement with
the PAGI mass spectrum slope (around $-2$) is a bit worse but still acceptably
good. One reason why the observed mass spectrum is slightly less steep than the
prediction of PAGI may be that the three most massive clumps include very
dense regions that are not resolved completely. With higher resolution,
these three clumps could perhaps split into smaller objects. At the
low-mass end, the observed CMF includes four clumps with masses substantially
lower than $m_\mathrm{CMF}$, whereas the PAGI mass function has a low-mass
cutoff. One reason for this may be incompleteness. We are unable to calculate
the completeness limit without making further assumptions about the internal
density and velocity structure of our clumps. Since our mass resolution
of $\sim 0.6$~M$_\odot$ is, however, substantially lower than the
maximum of the CMF at $\sim 10$~M$_\odot$, it is more probable that there really
exists a small number of low-mass clumps that cannot be explained by the
PAGI dispersion relation. They may be result of non-linear processes or
non-gravitational fragmentation.

\subsection{Spatial distribution of clumps}

\begin{figure}
\centering
\includegraphics[width=\columnwidth]{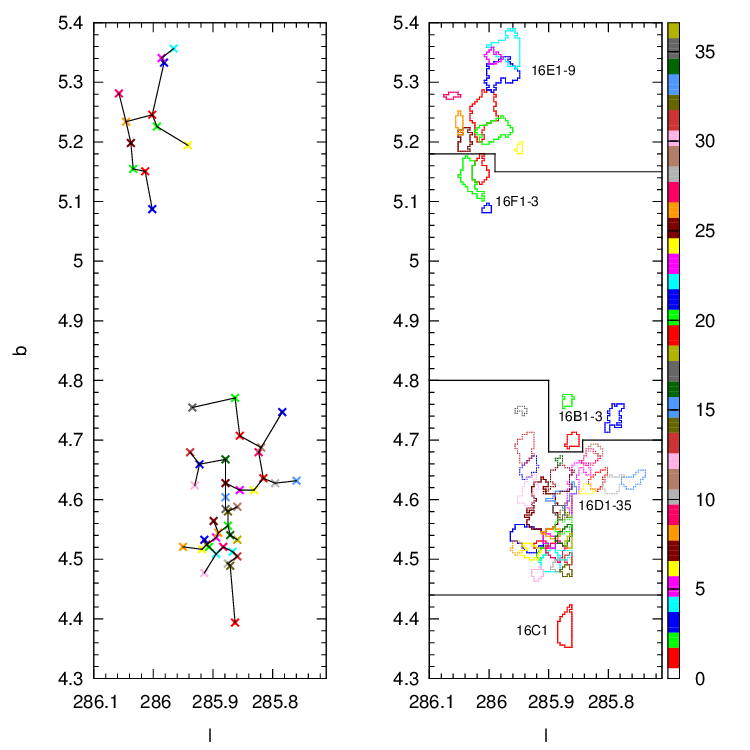}
\caption{Left: minimum spanning trees showing projected distances between clumps.
The two trees are constructed separately in the upper and in the lower observed
region. Each tree node (denoted by the $\times$ symbol) is at the position of the pixel
with the highest brightness temperatures in a given clump ($T_\mathrm{peak}$).
Right: outlines of the $^{13}$CO clumps projected onto the
$lb$-plane. Colors and line types correspond to those in
Figures~\ref{dendro1}--\ref{clumps_ex} and \ref{clumps1}--\ref{clumps5}.
}
\label{fig:MST}
\end{figure}

We estimate the typical separation of clumps, $d_\mathrm{MST}$, from the average
length of the edges of the minimum spanning tree (MST, see
Figure~\ref{fig:MST}). Each clump is represented by the point at the position of
the brightest pixel of the clump ($l_\mathrm{peak}$, $b_\mathrm{peak}$). In each
of the two observed regions, all the points are connected with lines -- tree
edges -- in such a way that the sum of all edges lengths is minimal. Taking into
account the distance to the Carina Flare, $2.6$~kpc, the average edge length in
the $lb$-plane corresponds to $\sim 1.4$~pc with a standard deviation of $\sim
1$~pc. Since we do not have information about the positions of clumps in the radial
direction, we assume that their separations in this direction are similar to
those in the other two directions, and multiply the average MST edge length in
the $lb$-plane by factor $\sqrt{3/2}$. We obtain
\begin{equation}
d_\mathrm{MST} = (1.7 \pm 1.2)\ \mathrm{pc} \ .
\end{equation}

\subsection{Clump relative velocities}

\begin{figure}
\centering
\includegraphics[width=\columnwidth]{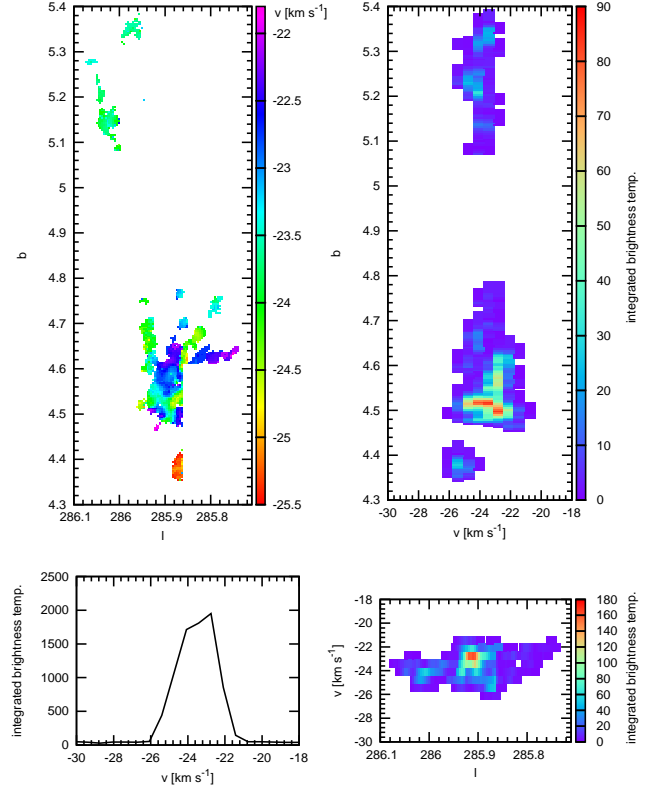}
\caption{Distribution of the velocity in the $^{13}$CO APEX data. Top left:
average velocity taken over all velocity channels weighted by $T_\mathrm{b}$. Top right:
$T_\mathrm{b}$ integrated along the $l$ coordinate shown in the $bv$-plane. Bottom
right: $T_\mathrm{b}$ integrated along the $b$ coordinate shown in the $lv$-plane.
Bottom left: the line profile integrated over both spatial coordinates.
}
\label{fig:vel}
\end{figure}

Figure~\ref{fig:vel} shows the radial velocity distribution of clumps found in
the $^{13}$CO data. All clumps are distributed across only seven velocity
channels, i.e. the maximum relative radial velocity of clumps is
$4.6$~km\,s$^{-1}$. In some regions, the velocity varies by $\sim
2$~km\,s$^{-1}$ on scales smaller than several pc of the projected distance, and
there is no obvious global gradient in any direction. This velocity variation is
much higher than would be expected from the stretching of the shell surface due
to the shell expansion. For instance, on a scale of $5$~pc the relative velocity
of clumps due to shell stretching is $\sim 0.3$~km\,s$^{-1}$ assuming the shell
radius and expansion velocity to be $150$~pc and $10$~km\,s$^{-1}$,
respectively, \citep{1999PASJ...51..751F}. It may be, however, that the clump
relative velocities are caused by their mutual gravitational attraction. For
instance, if we estimate the dynamical velocity for Cloud 16D, we get
$v_\mathrm{dyn,16D} \equiv (GM_\mathrm{16D}/R_\mathrm{16D})^{1/2} \sim
1.2$~km\,s$^{-1}$ using $M_\mathrm{16D} \sim 1.8\times 10^{3}$~M$_\odot$ and
$R_\mathrm{16D} \sim 5$~pc. Another source of clump-clump relative velocities
may be local processes like star formation and related feedback. Our spatial
resolution, however, does not allow us to see these small scale processes.

\subsection{Origin of clumps: PAGI}
\label{sec:pagi}

The Pressure Assisted Gravitational Instability (PAGI) is described in
\citet{2009MNRAS.398.1537D}, \citet{2010MNRAS.407.1963W},
and \cite{2011MNRAS.411.2230D}. For PAGI,
it is assumed that a thick shell expands into a relatively rarefied ambient
medium with non-zero pressure, $P_{_{\rm EXT}}$, and that the same pressure acts
also on the inner wall of the shell. The shell fragmentation is controlled by
the interplay between the gravitational force in the plane of the shell, the
thermal pressure inside the shell and the external pressure, $P_{_{\rm EXT}}$.
PAGI can be described by the following dispersion relation:

\begin{eqnarray}
\omega^2   \propto 
\frac{3 G\Sigma l}{4R}\Gamma(\beta)+
\frac{10P_{_{\rm EXT}}c_{_{\rm S}}^2 l^2}{3\pi^2
R^2(2P_{_{\rm EXT}}+\pi G\Sigma^2)}
-\frac{5 c_{_{\rm S}}^2 l^2}{2\pi^2R^2}\,,
\label{eq:PAGI:ms}
\end{eqnarray}
where $\omega$ is the growth rate of a perturbation with the dimensionless
wavenumber $l$. $R$, $\Sigma$ and $c_{_{\rm S}}$ are the shell
radius, surface density and internal speed of sound.
The geometrical factor $\Gamma$ results from the
approximation of a fragment by a uniform oblate spheroid and is given by
\begin{equation}
\Gamma(\beta) = \frac{\cos^{-1}\beta}{(1-\beta^2)^{3/2}} -
\frac{\beta}{1-\beta^2}
\end{equation}
where
\begin{equation}
\beta \equiv \frac{z}{r} = \frac{\Sigma c_{_{\rm S}}^2}{2P_{_{\rm EXT}}+\pi G
\Sigma^2} \frac{l}{\pi R}
\end{equation}
is the fragment aspect ratio (see \citealt{2010MNRAS.407.1963W} for details). In
Equation (\ref{eq:PAGI:ms}), we do not show
other terms dependent on the expansion velocity of the shell $V$, since the
contribution of these terms is small compared to those from terms describing
effects of the self-gravity, the external pressure, and the internal pressure,
i.~e., the first, second and third term of (\ref{eq:PAGI:ms}) respectively.
We also assume that the average molecular weight $\mu$ is 2.35 a.m.u. 

This dispersion relation produces a mass spectrum of fragments as
\begin{equation}
\frac{\mathrm{d}N}{\mathrm{d}m} \propto \omega \times  m^{-2},
\end{equation} 
where $m$ is the mass of an individual fragment and $\mathrm{d}N$ is the number of
fragments in the mass interval $(m, m+\mathrm{d}m)$.  A more detailed derivation of the
mass spectrum is given in  \citet{2011MNRAS.411.2230D}.

\begin{figure*}
\centering
\includegraphics[width=\textwidth]{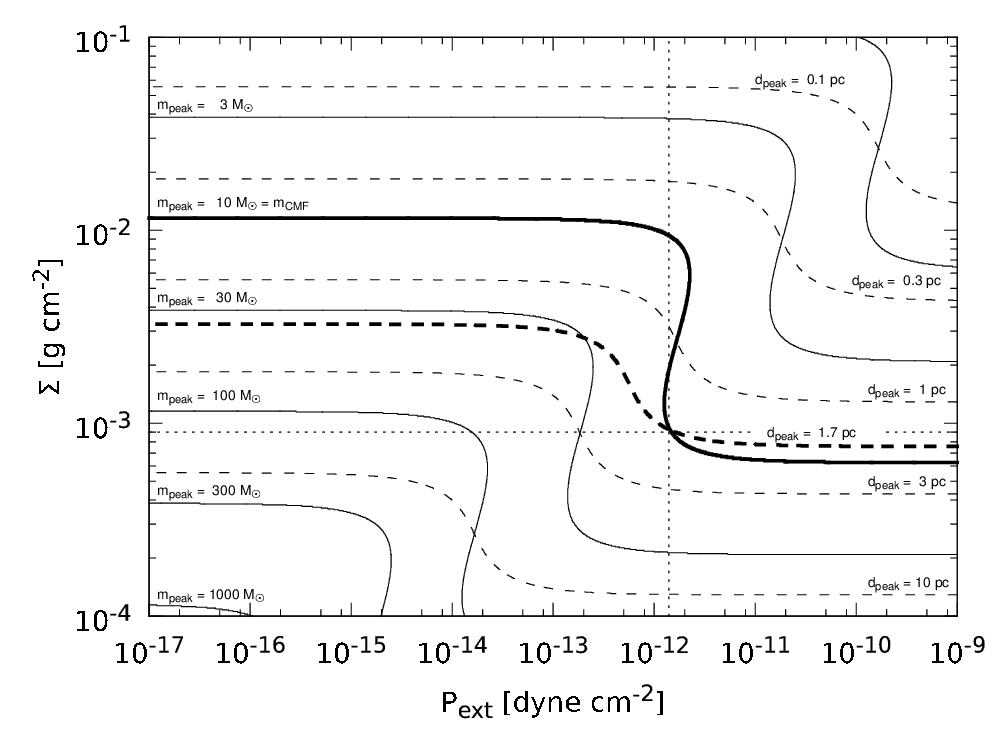}
\caption{Maximum of the mass spectrum, $m_\mathrm{peak}$, (solid isolines)
and the typical fragment separation, $d_\mathrm{peak}$, (dashed isolines) given
by PAGI as a function of the shell surface density $\Sigma$ and the external
pressure $P_{_{\rm EXT}}$. The thick lines show the maximum of the CMF,
$m_\mathrm{CMF}$, and typical separation of clumps, $d_\mathrm{MST}$, obtained
from our $^{13}$CO APEX observations. The thin dotted lines show the
$m_\mathrm{CMF}$ -- $d_\mathrm{MST}$ intersection at $\Sigma = 9\times 10^{-4}$~g\,cm$^{-2}$ and
$P_{_{\rm EXT}} = 1.4\times 10^{-12}$~dyne\,cm$^{-2}$.}
\label{fig:mdpeak}
\end{figure*}

The PAGI mass spectrum is in general a function of the fragment mass, $m$, the
surface density of the shell, $\Sigma$, the external pressure, $P_{_{\rm EXT}}$,
and the shell temperature, $T$. Solid isolines in Figure~\ref{fig:mdpeak} show
the maximum of the mass spectrum $m_\mathrm{peak}$ as a function of $\Sigma$ and
$P_{_{\rm EXT}}$. In line with previous works \citep{1999PASJ...51..751F,
2008PASJ...60.1297D}, we assume that the average temperature in clumps (and
hence in the shell) is $T = 15$~K. This temperature is lower than the
minimum we obtain from $^{13}$CO(J=2--1) observations in the densest regions
(above $20$~K) but it is probable that the temperature in these regions is
higher than average. In the derivation of PAGI, it is assumed that the mass of
each fragment originates from a circular area with a diameter $d = 2\pi R/l$.
The typical separation among neighbouring fragments is $d_\mathrm{peak} =
(4m_\mathrm{peak}/\pi\Sigma)^{1/2}$, shown by dashed isolines in
Figure~\ref{fig:mdpeak}. An intersection between an $m_\mathrm{peak}$ isoline
and a $d_\mathrm{peak}$ isoline gives $\Sigma$ and $P_{_{\rm EXT}}$ of the shell
which forms fragments with typical mass $m_\mathrm{peak}$ and typical separation
$d_\mathrm{peak}$.

Note that the intersection exists only for certain combinations of
$m_\mathrm{peak}$ and $d_\mathrm{peak}$. For instance, the $m_\mathrm{peak} =
3$~M$_\odot$ isoline never intersects with the $d_\mathrm{peak} = 1$~pc. Such
fragments are inconsistent with the PAGI theory because their surface
densities are too low for fragmentation. Similarly, the $m_\mathrm{peak} =
30$~M$_\odot$ isoline also never intersects with the $d_\mathrm{peak} = 1$~pc
but in this case, the surface density of such fragments is too high
and they would break up into smaller objects. Therefore, this analysis provides
quite a strong test if fragments with given masses and separations can be
created by PAGI or not.

Taking the properties of clumps obtained from the $^{13}$CO APEX observations
($m_\mathrm{peak} = m_\mathrm{CMF} = 10$~M$_\odot$ and $d_\mathrm{peak} =
d_\mathrm{MST} = 1.7$~pc), we see that the corresponding isolines
intersect at $\Sigma = 9\times 10^{-4}$~g\,cm$^{-2}$ and $P_{_{\rm EXT}} =
1.4\times 10^{-12}$~dyne\,cm$^{-2}$ (the isolines are shown by the thick lines
and resulting $\Sigma$ and $P_{_{\rm EXT}}$ by thin dotted lines in
Figure~\ref{fig:mdpeak}). Considering the uncertainties in
determinations of $m_\mathrm{CMF}$ and $d_\mathrm{MST}$, the clump properties
would be also consistent with $P_{_{\rm EXT}} > 1.4\times
10^{-12}$~dyne\,cm$^{-2}$ where the isolines are parallel, close to each other. 

The obtained value of surface density is in an excellent agreement with the
average surface density of the Carina Flare supershell determined in previous
works ($\Sigma = M/(4\pi R^2)\sim 7\times 10^{-4}$~g\,cm$^{-2}$ with the total
mass of the shell $M \sim 10^6$~M$_\odot$ and its radius $R \sim 150$~pc; see
\citealt{2008MNRAS.387...31D}). The determined external pressure is somewhat
higher that the thermal pressure thought to be present at this latitude
(corresponding to $z \sim 250$~pc) above the galactic plane. The HI disc,
however, cannot be supported against gravity by thermal pressure
alone and other means of support have been suggested, e.g., turbulence
\citep{1990ARA&A..28..215D}.

\section{Conclusions}
\label{sec:concl}

We find that fragmentation in Cloud~16 of the Carina Flare has proceeded
to much smaller scales than were visible in the NANTEN observations of
\cite{2008PASJ...60.1297D}. It is probable, however, that there are structures
present in this region that our APEX observations cannot resolve. 
The superior angular and velocity resolution of ALMA may allow the
limits of the fragmentation process to be observed and accretion flows feeding
clumps to be disentangled.

Our theoretical work \citep[PAGI]{2009MNRAS.398.1537D, 2010MNRAS.407.1963W,
2011MNRAS.411.2230D} predicts that, in the early linear stages of shell
fragmentation, the mass and the wavelength at which the shell preferentially
fragments depends strongly on the confining pressure. Here we show that the CMF
and the typical clump separation in the observed part of the Cloud~16 are
consistent with the fragmentation caused by PAGI. The surface density of the
fragmenting layer, determined from the observed clump properties, is $\Sigma =
9\times 10^{-4}$~g\,cm$^{-2}$, a value in good agreement with the
estimated average surface density of the Carina Flare supershell. The clump
properties also suggest that the pressure confining the shell should be
$\sim 1.4\times 10^{-12}$~dyne\,cm$^{-2}$ or higher, we are unable, however, to
exclude the possibility that the shell is in more advanced stages of the
fragmentation for which the PAGI dispersion relation is not accurate.

Since the extent of the Carina Flare above the Galactic Plane is several scale
heights, the pressure encountered by the top part of the shell may be several
times smaller than that confining the sides. We intend to perform observations
of the molecular cloud G287.60+8.00 (Cloud 74 of \citealt{2008PASJ...60.1297D}),
located at the greatest z-extent of the supershell, to see if the effect
of the lower pressure environment on fragment size can be detected there. 

\begin{acknowledgements}
We thank the anonymous referee and Steve Shore for many suggestions which
significantly improved the paper. We thank Michael Dumke for his assistance
during APEX observations. This study was supported by the Institutional Research
Plan AV0Z10030501 of the Academy of Sciences of the Czech Republic and project
LC06014 Centre for Theoretical Astrophysics of the Ministry of Education, Youth
and Sports of the Czech Republic. VS acknowledges support from Doctoral grant of
the Czech Science Foundation No.~205/09/H033 and a grant of the Charles
University in Prague No.~SVV~261301.

\end{acknowledgements}

\bibliography{myrefs}

\begin{appendix}

\section{The DENDROFIND algorithm: searching the clump tree}
\label{ap:df}

\begin{figure*}
\includegraphics[width=\textwidth]{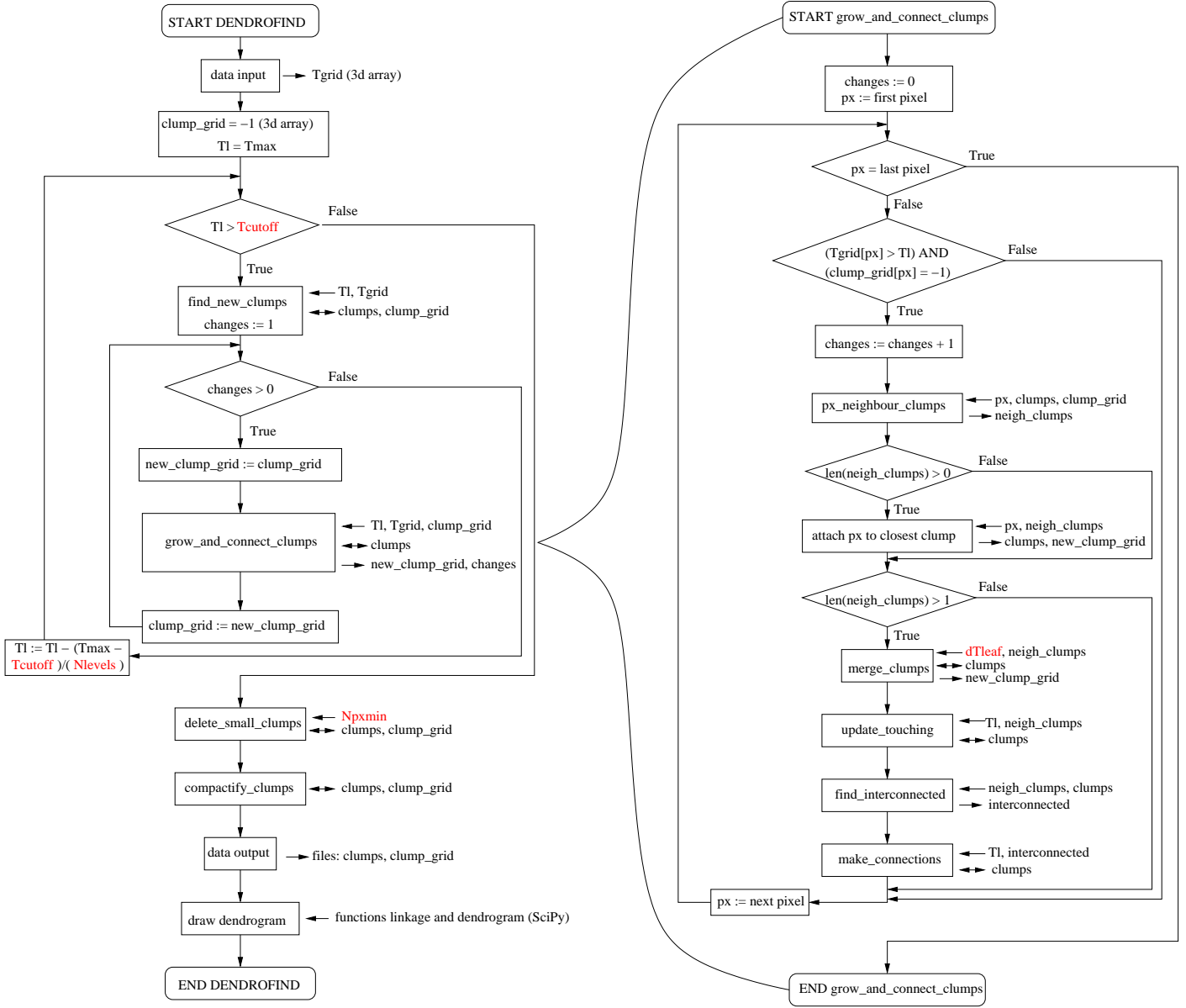}
\caption{The DENDROFIND flow chart. The input parameters are printed in red
color. Left and right arrows with variables next to the boxes with algorithm
functions show input and output to/from this function, respectively. Double
head arrows denote variables used for both input and output.}
\label{fig:flowchart}
\end{figure*}

The DENDROFIND package can be found at \mbox{\tt http://galaxy.asu.cas.cz/\textasciitilde
richard/dendrofind}. It consists of several python scripts and uses the
libraries NumPy \citep{numpy} and SciPy \citep{scipy}.
Figure~\ref{fig:flowchart} shows a flow-chart of the algorithm, the input
parameters are shown by Table~\ref{tab:df:input} and the essential data
structures are listed in Table~\ref{tab:df:variables}.

\subsection{Description of the algorithm}

The four input parameters \texttt{Nlevels}, \texttt{Npxmin},
\texttt{dTleaf} and \texttt{Tcutoff} are described in section
\S\ref{sec:dendrofind}. Setting \texttt{Nlevels} is straightforward: it
should have a value such that increases do not lead to different
results. The value of \texttt{Npxmin} depends on the resolution of the
instrument (or the numerical method) and it should correspond to the
number of pixels of the smallest resolvable structure. Finding values of
\texttt{dTleaf} and \texttt{Tcutoff} is not so obvious. Fortunately, the
dependence of results on these two parameters is quite intuitive. Decreasing the
value of \texttt{dTleaf} leads to higher numbers of identified structures (as
it is easier for local maxima to become independent clumps); and decreasing the
value of \texttt{Tcutoff} leads to higher fractions of the datacube 
assigned to clumps (and in most cases also to higher numbers of clumps). In many
cases, reasonable values for both parameters can be set to some multiple of
the noise dispersion, e.~g., \texttt{dTleaf} = \texttt{Tcutoff} =
$3\sigma_\mathrm{noise}$ as done in this work.

The most important data structures for the algorithm are the two 3D arrays
\texttt{Tgrid} and \texttt{clump\_grid} and the list \texttt{clumps}. The
\texttt{Tgrid} array is a copy of the PPV data cube including the brightness
temperature $T_\mathrm{b}$ for each pixel \texttt{px}. The \texttt{clump\_grid}
array has the same dimensions as \texttt{Tgrid}, and it includes, for each pixel
\texttt{px}, the ID number of a clump to which pixel \texttt{px} is assigned
($-1$ means that the pixel is not assigned to any clump). The list
\texttt{clumps} consists of all the identified clumps represented as data
objects that include all necessary information. In particular, each
object contains the clump ID number, the position and value of the clump peak
temperature, the list \texttt{pixels} of pixels that are assigned to the
clump, the list \texttt{touching} of clumps that directly touch the clump, and
the list \texttt{connected} of clumps that are connected (even through other
clumps) to the clump. The lists \texttt{touching} and \texttt{connected} also
record information about the temperature at which the clump is touching or
connected to other clumps. Another important piece of information given is
the list \texttt{neigh} that includes ID numbers of clumps 
neighbouring the processed pixel. In a recent version of the code, neighbours
of pixel \texttt{px} are defined as pixels which share with \texttt{px} a face,
i.e. each pixel has six neighbours. A simple single modification of the
code, however, can change the behaviour so that any other pixels (for instance
diagonal ones) are also taken into account as neighbours.

The algorithm starts by reading the data from the fits file and creating the
array \texttt{Tgrid} (see Figure~\ref{fig:flowchart}). Then the array
\texttt{clump\_grid} is created and the value -1 is assigned to all its pixels.
Next, the maximum brightness temperature \texttt{Tmax} is determined and
the main loop of the algorithm is entered. It executes \texttt{Nlevels} iterations
decreasing the variable \texttt{Tl} from \texttt{Tmax} to \texttt{Tcutoff}.

In each iteration of the main loop, all pixels \texttt{px} with
\texttt{Tgrid(px) > Tl} and \texttt{clump\_grid(px) = -1} are processed in
several sweeps through the \texttt{Tgrid} array. In the first sweep, a new clump
is created for each processed pixel which is a local maximum (i.~e.,
\texttt{Tgrid(px)} is greater than \texttt{Tgrid} of all \texttt{px}
neighbours). An unused ID number is found and assigned to
\texttt{clump\_grid(px)} and a new clump object is created and added to the list
\texttt{clumps}. In subsequent sweeps, processed pixels which are not local maxima are
assigned to existing clumps and connections among clumps are found (see
description of function \texttt{grow\_and\_connect\_clumps} below). To
make the procedure independent of the order in which the pixels are processed in
one sweep, information about pixel assignments to existing clumps is written
into a copy of the \texttt{clump\_grid} array \texttt{new\_clump\_grid}. At the
end of the sweep, \texttt{new\_clump\_grid} is copied back to
\texttt{clump\_grid} and another sweep is made until no assignment occurs (this
is controlled by the variable \texttt{changes}).

The \texttt{grow\_and\_connect\_clumps} function consists of several steps
executed for each processed pixel (i.~e., one with \texttt{Tgrid(px) > Tl} and
\texttt{clump\_grid(px) = -1}). At first, all clumps to which neighbour pixels
of the processed pixel belong are identified and stored in the list
\texttt{neigh}. If \texttt{neigh} includes at least one clump, the processed
pixel \texttt{px} is assigned to the clump from \texttt{neigh} whose peak is
closest to \texttt{px}. Further, if the \texttt{neigh} list includes more than
one clump, four steps are executed. Firstly, clumps from the \texttt{neigh} list
with \texttt{Tpeak - Tl < dTleaf}, where \texttt{Tpeak} is the peak temperature
of the clump, are merged with the clump from the \texttt{neigh} list with the
highest \texttt{Tpeak}. Then, the lists \texttt{touching} of all clumps in the
\texttt{neigh} list are updated so that each clump includes all other clumps
from the \texttt{neigh} list on its respective \texttt{touching} list. Next,
the \texttt{interconnected} list is found of all clumps that can be reached from the
pixel \texttt{px} by moving only through neighbour pixels with temperatures
greater than \texttt{Tl}. Finally, the \texttt{connected} lists
of all clumps are updated so that each clump includes all the other clumps from
the \texttt{interconnected} list on its \texttt{connected} list.

At the end, after the \texttt{Tcutoff} limit is reached, clumps with fewer than \texttt{Npxmin} pixels are deleted and all clumps are renumbered
so that their IDs are consecutive. Then, the array \texttt{clump\_grid} and the
list \texttt{clumps} are written into files. Finally, the SciPy functions
\texttt{linkage} and \texttt{dendrogram} are used to create the dendrogram. The
function \texttt{linkage} organizes clumps into hierarchical clusters based
on their \texttt{connected} lists. The function \texttt{dendrogram} actually
plots the dendrogram finding the exact positions of the dendrogram lines in
the figure.

\begin{table}
\begin{tabular}{l|l}
input param. & description \\
\hline
\texttt{Nlevels} & number of temperature levels \\
\texttt{Npxmin}  & min. number of pixels of a clump \\
\texttt{dTleaf}  & min. length (in temperature) of a dendrogram leaf\\
\texttt{Tcutoff} & min. $T_\mathrm{b}$ considered for assignment to clumps \\
\end{tabular}
\caption{List of input parameters.}
\label{tab:df:input}
\end{table}

\begin{table}
\begin{tabular}{l|l}
data structure & description \\
\hline
\texttt{Tgrid}       & 3D array (PPV) of the brightness temperature \\
\texttt{clump\_grid} & 3D array (PPV) of clump ID numbers \\
\texttt{clumps}      & list of clump objects \\
\texttt{neigh}       & IDs of clumps neighbouring with a processed px. \\
\texttt{len(neigh)}  & length of the neigh array \\
\texttt{Tl}          & actual brightness temp. level in the main loop\\
\texttt{px}          & index of the processed pixel (in the code i,j,k) \\
\end{tabular}
\caption{List of variables important for the description of the algorithm.}
\label{tab:df:variables}
\end{table}

\subsection{Sensitivity to noise}

\begin{figure}
\centering
\includegraphics[width=\columnwidth]{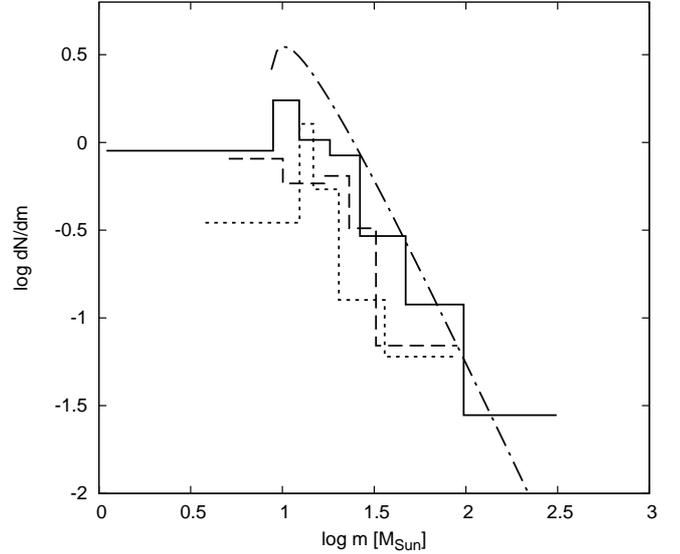}
\caption{Dependence of the clump mass spectrum on the noise in the datacube of
the brightness temperature. The solid line shows the original mass spectrum
(with $\sigma_T = 0.19$~K), the dashed and dotted lines show spectra obtained
from datacubes where the noise was artificially increased by factor 2 and 3,
respectively. The dash-dotted line shows the PAGI mass spectrum for the same
parameters as in Figure~\ref{fig:ms}.
}
\label{fig:dftestn}
\end{figure}

To estimate how sensitive the DENDROFIND algorithm is to the level of
noise present in the data, we calculate the $^{13}$CO clump mass spectrum for
three different levels of noise (see Figure~\ref{fig:dftestn}). The solid line
shows the original CMF with the noise level $\sigma_T \simeq 0.19$~K, the dashed
and dotted lines show CMFs obtained for $\sigma_T = 0.38$~K and $\sigma_T =
0.57$~K. The resulting spectra differ mainly in the number of objects (higher
$\sigma_T$ leads to lower numbers of identified clumps), but the slope of the
spectrum and its maximum ($m_\mathrm{peak}$) remain approximately the same.
There are two reasons for the dependence of the CMF on the noise level. Firstly,
the noise modifies the brightness temperature datacube and results in
slightly different masses, positions and interconnecting brightness temperatures
of clumps. The second, more important reason is that the noise level determines
values of parameters \texttt{Tcutoff} and \texttt{dTleaf} which are both set to
$3\sigma_T$. We conclude that the CMF slope and peak mass are not significantly
changed if the noise is increased by up to a factor of 3.

\subsection{Sensitivity to the velocity resolution}

\begin{figure}
\centering
\includegraphics[width=\columnwidth]{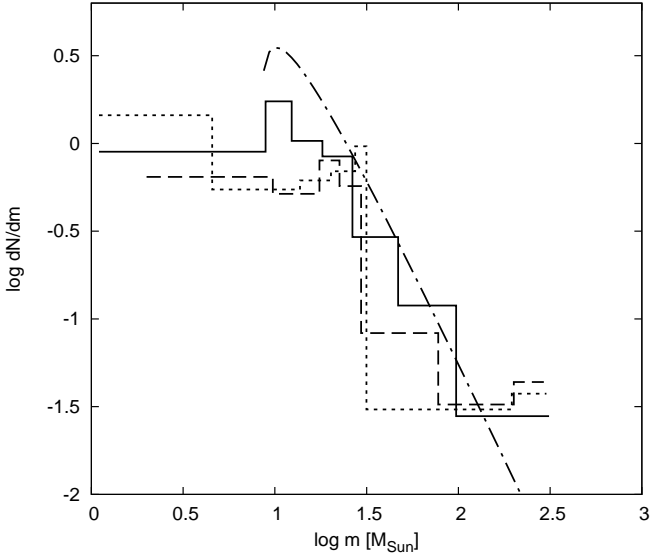}
\caption{Dependence of the clump mass spectrum on the velocity resolution in the
datacube. The solid line shows the original mass spectrum ($dv =
0.66$~km\,s$^{-1}$), the dashed and dotted lines show spectra obtained if 2 and
3 velocity channels, respectively, are added together. The dash-dotted line
shows the PAGI mass spectrum for the same parameters as in Figure~\ref{fig:ms}. }
\label{fig:dftestv}
\end{figure}

We also test how sensitive the DENDROFIND algorithm is to the velocity
resolution. We do it by comparing the CMF obtained from the original data to
CMFs resulting from datacubes where two or three velocity channels were added
together. In Figure~\ref{fig:dftestv}, the solid line shows the original CMF
with the velocity channel width $dv = 0.66$~km\,s$^{-1}$, the dashed and dotted
lines show CMFs from datasets in which the velocity resolution was reduced by
factor of 2 and 3, respectively. The CMF with twice reduced velocity resolution 
still shows high mass end that is consistent with a power law of slope
$\sim -2$, and a peak at approximately $\sim 20$~M$_\odot$. The CMF
with $dv = 2$~km\,s$^{-1}$, however, does not agree with a power-law distribution at
all.

The reason for this behaviour is that the velocity resolution of our APEX
observation is already quite low, and most clumps have FWHM$_v$ smaller than $dv
= 1.34$~km\,s$^{-1}$ (the original velocity resolution reduced by 2). Therefore,
degrading the velocity resolution further results in a drop of the noise in the
datacube (and consequently decrease of parameters \texttt{Tcutoff} and
\texttt{dTleaf}). The signal-to-noise ratio, however, is not improved. Therefore,
as the velocity resolution decreases, the algorithm tends to find more and more
artificial structures.

Since most of clumps span only a few velocity channels, and it rarely
happens that the clumps projected onto the $lb$-plane overlap (see
Figure~\ref{fig:MST}, right panel), it is quite improbable that one real object is
misidentified as several objects due to internal velocity flows. We cannot
exclude the possibility, however, that some clumps could split into several less
massive objects at higher velocity resolution.

\end{appendix}

\Online
\begin{appendix}
\section{Clump properties and individual velocity channel maps}
\label{ap:figs}

\begin{figure*}
\centering
\includegraphics{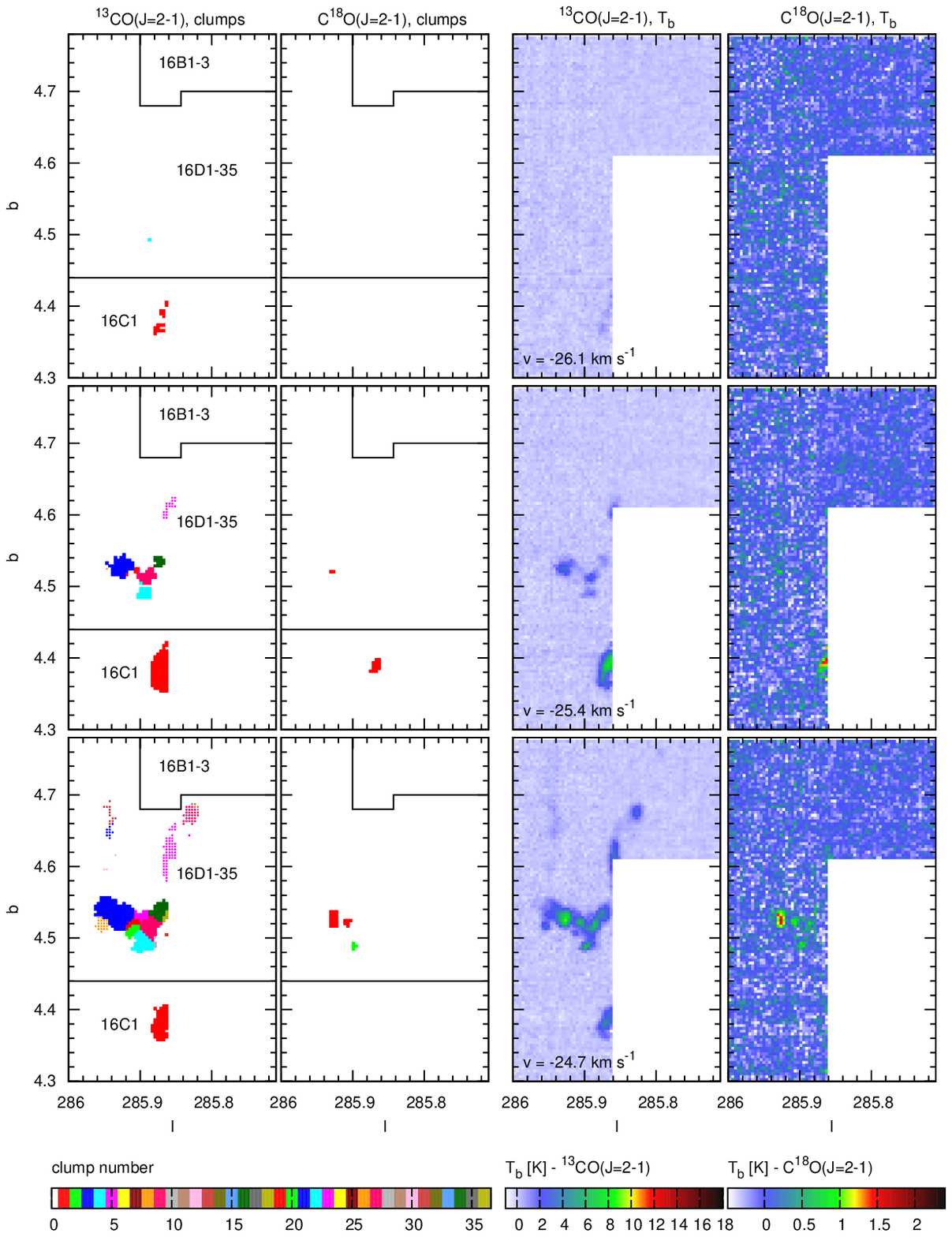}
\caption{Clumps identified in the lower region (left two panels) and
corresponding brightness temperature (right two panels) in the first three
velocity channels. Filled regions show clumps 1-18, dotted regions show clumps
19-35. Colors of clumps correspond to colors of dendrogram leaves in
Figure~\ref{dendro1}.}
\label{clumps1}
\end{figure*}

\begin{figure*}
\centering
\includegraphics{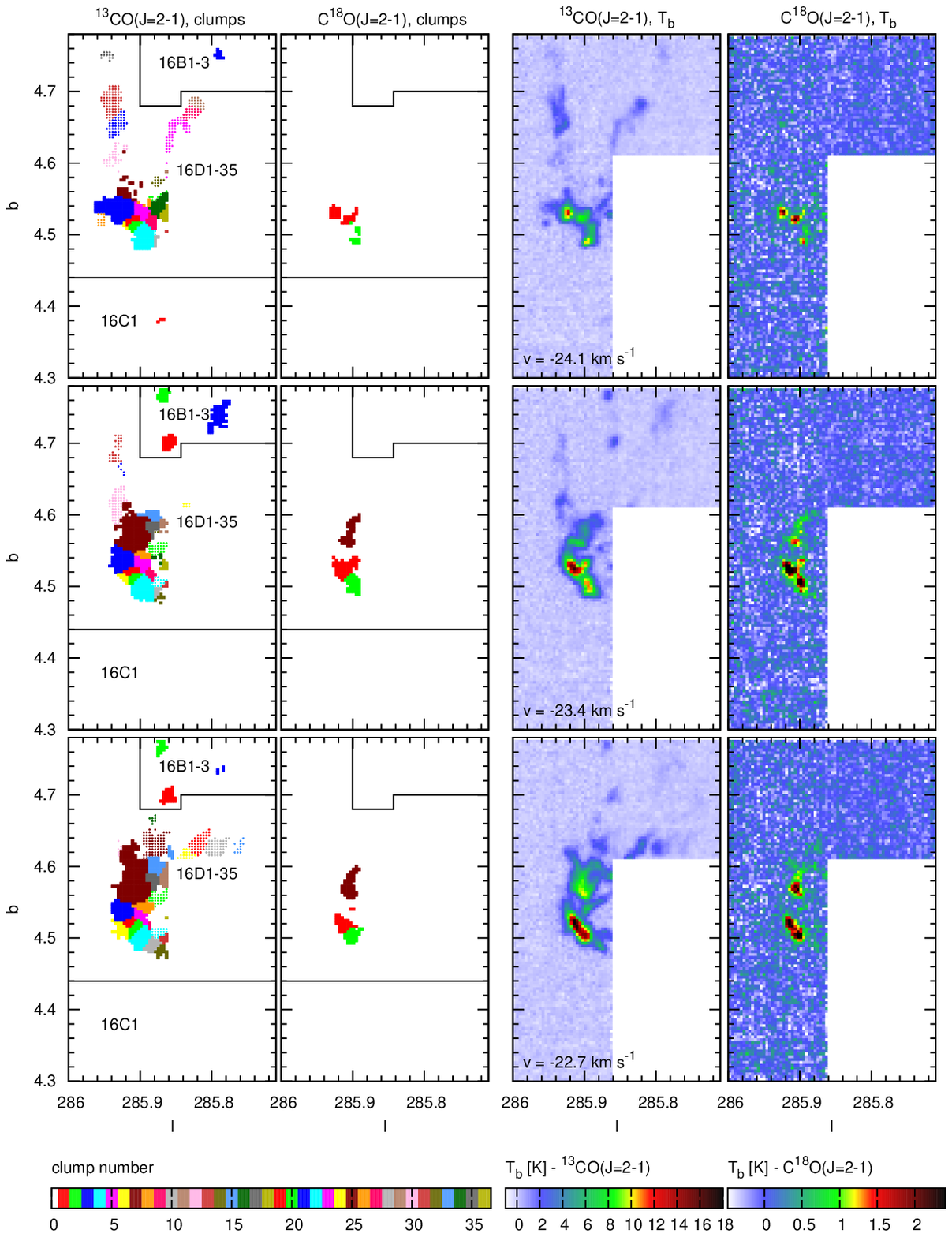}
\caption{Clumps identified in the lower region (left two panels) and
corresponding brightness temperature (right two panels) in the second three
velocity channels. Filled regions show clumps 1-18, dotted regions show clumps
19-35. Colors of clumps correspond to colors of dendrogram leaves in
Figure~\ref{dendro1}.}
\label{clumps2}
\end{figure*}

\begin{figure*}
\centering
\includegraphics{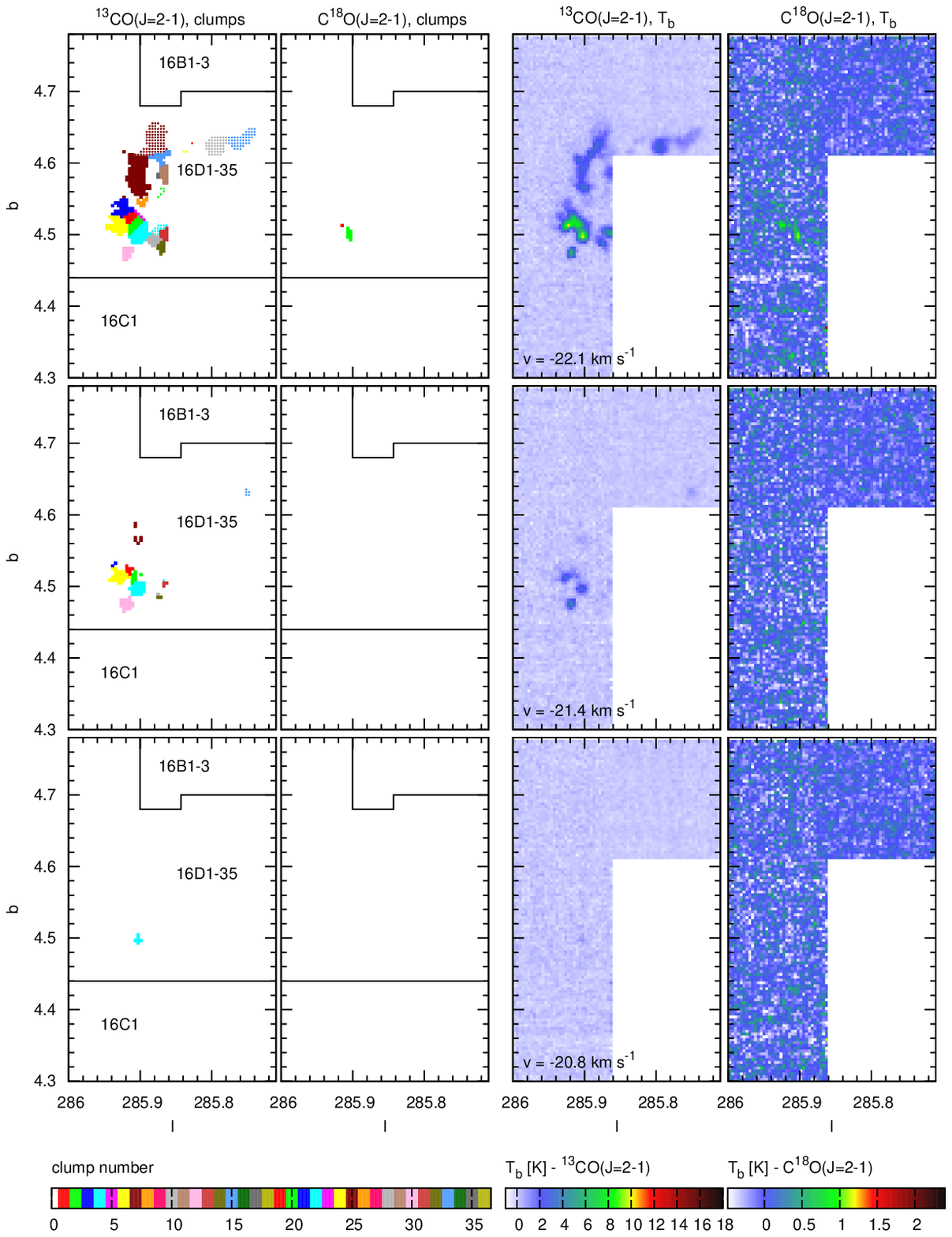}
\caption{Clumps identified in the lower region (left two panels) and
corresponding brightness temperature (right two panels) in the last three
velocity channels. Filled regions show clumps 1-18, dotted regions show clumps
19-35. Colors of clumps correspond to colors of dendrogram leaves in
Figure~\ref{dendro1}.}
\label{clumps3}
\end{figure*}

\begin{figure*}
\centering
\includegraphics{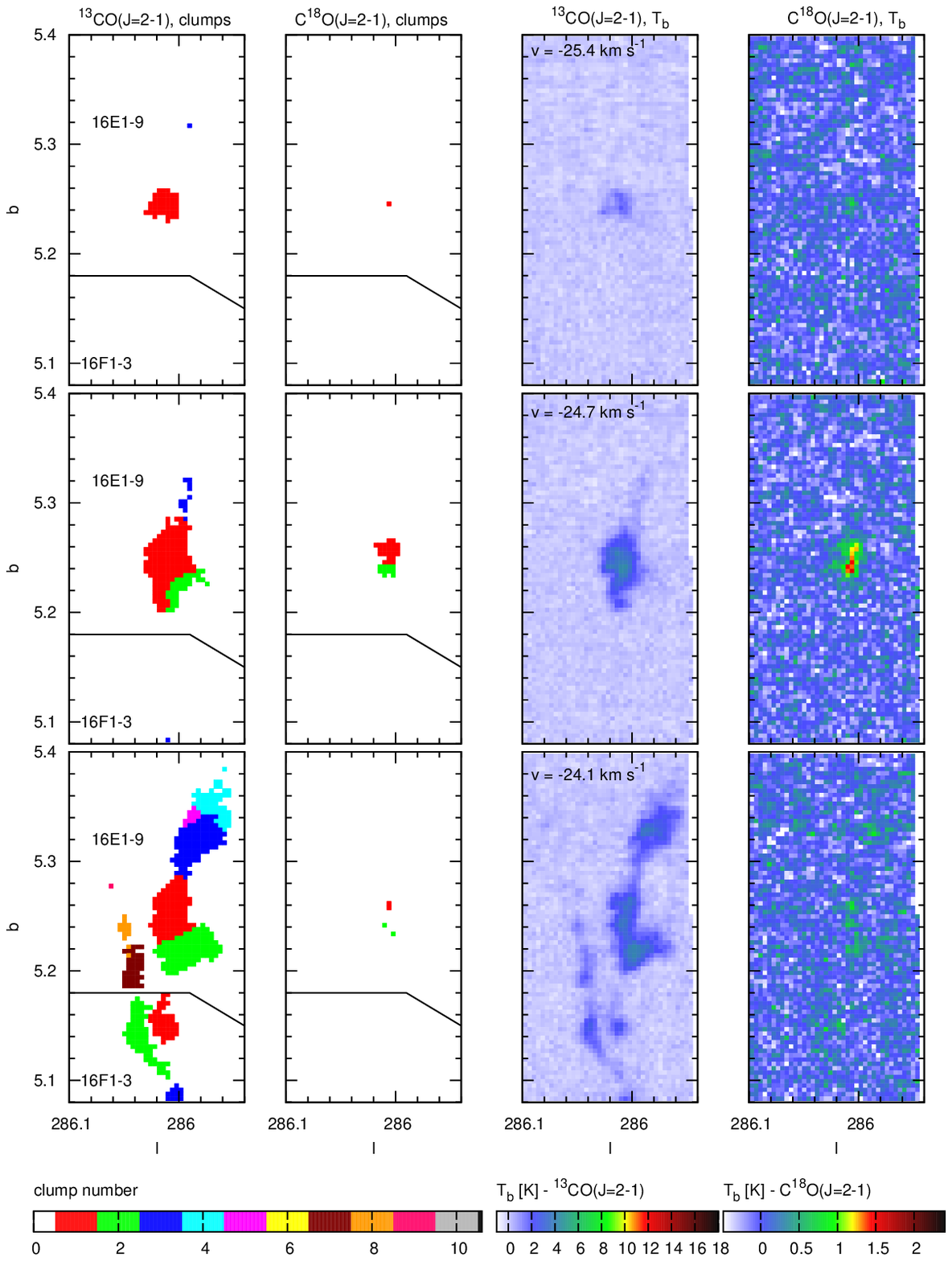}
\caption{Clumps identified in the upper region (left two panels) and
corresponding brightness temperature (right two panels) in the first three
velocity channels. Colors of clumps correspond to colors of dendrogram leaves in
Figure~\ref{dendro2}.}
\label{clumps4}
\end{figure*}

\begin{figure*}
\centering
\includegraphics{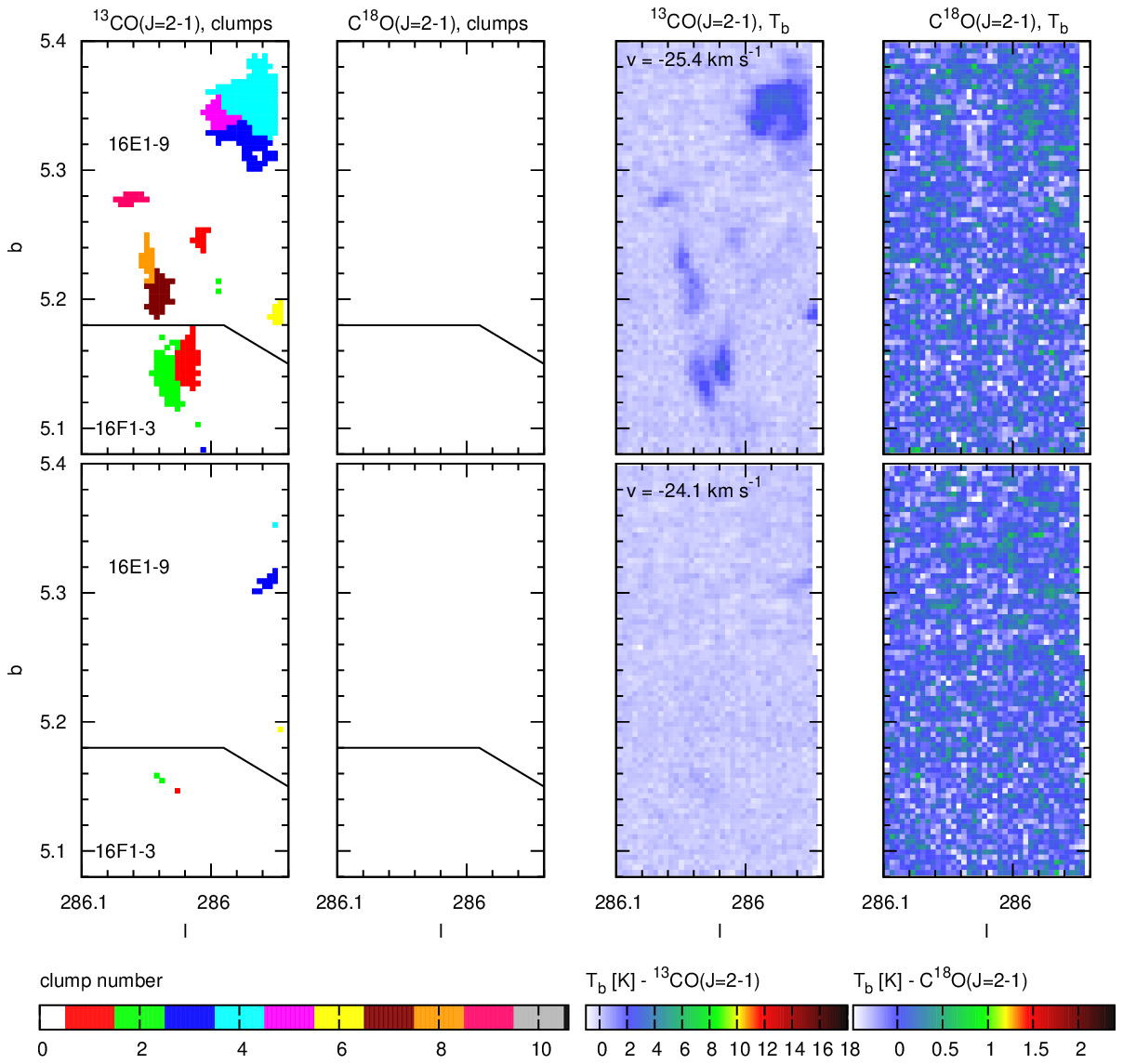}
\caption{Clumps identified in the upper region (left two panels) and
corresponding brightness temperature (right two panels) in the last two
velocity channels. Colors of clumps correspond to colors of dendrogram leaves in
Figure~\ref{dendro2}.}
\label{clumps5}
\end{figure*}

In this appendix we present full versions of
Tables~\ref{table:13CO_clumps_lower} and \ref{table:13CO_clumps_upper} and
individual velocity channel maps together with maps showing identified clumps
(see Figures~\ref{clumps1}--\ref{clumps5}).


\begin{table*}
  \centering
\begin{tabular}{lrrrrrrrrrrrrr}
  Label &
  $\frac{\mathrm{FWHM}_{x}}{\mathrm{pc}}$ &
  $\frac{\mathrm{FWHM}_{y}}{\mathrm{pc}}$ &
  $\frac{\mathrm{FWHM}_{v}}{\mathrm{km\;s^{-1}}}$ &
  $\frac{R}{\mathrm{pc}}$ &
  $\frac{T_{\mathrm{peak}}}{\mathrm{K}}$ &
  $\frac{l_\mathrm{peak}}{\mathrm{deg}}$ &
  $\frac{b_\mathrm{peak}}{\mathrm{deg}}$ &
  $\frac{v_\mathrm{peak}}{\mathrm{km\;s^{-1}}}$ &
  $\frac{m}{\mathrm{M}_\odot}$ &
  $\frac{\Delta m}{\mathrm{M}_\odot}$ &
  $\frac{m_{\mathrm{vir}}}{\mathrm{M}_\odot}$ &
  $\frac{\Delta m_{\mathrm{vir}}}{\mathrm{M}_\odot}$ &
  $\frac{m}{m_{\mathrm{vir}}}$
  \\
  \hline\hline
  16D1  &   0.59 &   0.37 &   1.98 &   0.4 &   18.3 & 285:54:51.7 & 4:31:16.6 & -22.75 &  144.6 & 0.33 &  330.7 &    221.4 &    0.437 \\
  16D2  &   0.51 &   0.51 &   1.91 &   0.5 &   17.4 & 285:54:37.4 & 4:31:02.3 & -22.75 &  119.6 & 0.37 &  315.2 &    218.9 &    0.379 \\
  16D3  &   1.08 &   0.90 &   2.07 &   1.1 &   15.5 & 285:55:05.9 & 4:31:45.2 & -23.41 &  312.2 & 0.79 &  895.8 &    574.0 &    0.348 \\
  16D4  &   0.73 &   0.89 &   2.16 &   0.8 &   14.7 & 285:53:54.6 & 4:30:19.3 & -22.75 &  292.8 & 0.68 &  722.7 &    445.1 &    0.405 \\
  16D5  &   0.58 &   0.57 &   1.48 &   0.5 &   10.9 & 285:53:54.4 & 4:31:59.2 & -23.41 &   75.0 & 0.37 &  223.6 &    200.6 &    0.336 \\
  16D6  &   0.64 &   0.57 &   1.56 &   0.6 &   10.2 & 285:55:20.3 & 4:30:48.1 & -22.08 &   52.0 & 0.40 &  295.9 &    252.3 &    0.176 \\
  16D7  &   1.09 &   1.88 &   1.28 &   1.4 &   10.0 & 285:54:08.5 & 4:33:39.2 & -22.75 &  307.7 & 0.85 &  435.0 &    452.7 &    0.707 \\
  16D8  &   0.68 &   0.44 &   1.10 &   0.5 &    6.8 & 285:53:40.0 & 4:32:27.7 & -22.75 &   26.2 & 0.29 &  115.3 &    138.6 &    0.228 \\
  16D9  &   0.64 &   0.67 &   1.52 &   0.7 &    6.2 & 285:53:11.7 & 4:31:01.9 & -24.74 &   49.1 & 0.40 &  298.7 &    261.0 &    0.164 \\
  16D10 &   0.51 &   0.50 &   1.43 &   0.5 &    6.1 & 285:52:43.3 & 4:29:21.8 & -22.75 &   26.7 & 0.30 &  194.1 &    179.8 &    0.138 \\
  16D11$\star$& 0.37 & 0.70 & 1.16 &   0.5 &    6.1 & 285:51:45.4 & 4:35:04.3 & -22.75 &   19.1 & 0.28 &  118.5 &    135.9 &    0.161 \\
  16D12 &   0.42 &   0.49 &   1.02 &   0.5 &    6.1 & 285:55:06.3 & 4:28:25.3 & -22.08 &   18.2 & 0.24 &   92.4 &    120.0 &    0.196 \\
  16D13$\star$& 0.38 & 0.42 & 1.36 &   0.4 &    5.9 & 285:51:46.1 & 4:30:04.4 & -22.08 &   13.2 & 0.21 &  124.0 &    120.9 &    0.106 \\
  16D14 &   0.42 &   0.46 &   1.28 &   0.4 &    5.2 & 285:52:29.1 & 4:29:07.5 & -22.75 &   10.0 & 0.23 &  130.4 &    135.2 &    0.077 \\
  16D15 &   0.84 &   0.80 &   1.12 &   0.7 &    5.1 & 285:52:56.7 & 4:36:01.7 & -22.75 &   32.4 & 0.35 &  170.4 &    202.4 &    0.190 \\
  16D16 &   0.61 &   0.79 &   1.22 &   0.6 &    4.7 & 285:52:28.6 & 4:32:13.1 & -24.74 &   30.6 & 0.36 &  178.9 &    194.9 &    0.171 \\
  16D17 &   0.56 &   0.58 &   1.04 &   0.5 &    4.5 & 285:52:56.9 & 4:34:50.3 & -22.75 &   20.6 & 0.26 &  100.3 &    127.9 &    0.206 \\
  16D18 &   0.33 &   0.48 &   1.20 &   0.4 &    4.3 & 285:51:45.8 & 4:31:44.4 & -24.08 &    9.4 & 0.20 &  100.9 &    111.2 &    0.094 \\
  16D19 &   0.69 &   0.86 &   0.68 &   0.7 &    4.1 & 285:49:07.8 & 4:37:55.0 & -22.75 &   15.7 & 0.26 &   57.6 &    112.7 &    0.272 \\
  16D20 &   0.75 &   0.57 &   1.24 &   0.6 &    4.0 & 285:52:42.8 & 4:33:10.3 & -22.75 &   18.1 & 0.33 &  174.9 &    187.6 &    0.103 \\
  16D21 &   0.74 &   0.91 &   0.86 &   0.7 &    3.8 & 285:55:33.6 & 4:39:21.9 & -24.08 &   20.1 & 0.31 &   95.6 &    148.5 &    0.211 \\
  16D22 &   0.61 &   0.44 &   1.40 &   0.5 &    3.7 & 285:52:14.6 & 4:30:33.1 & -22.75 &   16.3 & 0.28 &  167.9 &    159.8 &    0.097 \\
  16D23$\star$& 0.80 & 2.02 & 1.18 &   0.9 &    3.7 & 285:51:30.9 & 4:36:44.2 & -24.74 &   30.6 & 0.44 &  236.4 &    265.6 &    0.129 \\
  16D24$\star$& 0.55 & 0.43 & 0.90 &   0.4 &    3.7 & 285:50:05.1 & 4:36:43.9 & -22.75 &    8.4 & 0.20 &   66.0 &     97.6 &    0.127 \\
  16D25 &   0.92 &   1.16 &   1.02 &   1.0 &    3.5 & 285:52:56.5 & 4:37:27.3 & -22.08 &   44.8 & 0.49 &  190.8 &    248.1 &    0.235 \\
  16D26 &   0.44 &   0.59 &   1.01 &   0.5 &    3.4 & 285:57:14.6 & 4:31:02.8 & -24.74 &   10.4 & 0.23 &   90.3 &    118.6 &    0.116 \\
  16D27 &   0.60 &   0.65 &   1.01 &   0.6 &    3.2 & 285:49:36.0 & 4:40:32.2 & -24.74 &   15.4 & 0.30 &  114.3 &    150.6 &    0.135 \\
  16D28 &   0.78 &   0.74 &   1.02 &   0.7 &    3.0 & 285:47:56.4 & 4:37:26.1 & -22.08 &   28.0 & 0.38 &  141.7 &    184.6 &    0.198 \\
  16D29 &   0.55 &   0.45 &   0.87 &   0.5 &    2.8 & 285:49:21.7 & 4:41:00.7 & -24.08 &    7.4 & 0.20 &   64.9 &     99.4 &    0.114 \\
  16D30 &   0.70 &   1.32 &   1.12 &   0.8 &    2.8 & 285:56:02.5 & 4:37:13.6 & -23.41 &   21.2 & 0.39 &  196.2 &    233.7 &    0.108 \\
  16D31 &   0.71 &   1.35 &   1.01 &   0.9 &    2.7 & 285:56:30.7 & 4:40:33.4 & -24.08 &   26.1 & 0.41 &  170.8 &    223.5 &    0.153 \\
  16D32 &   0.41 &   0.42 &   0.83 &   0.4 &    1.5 & 285:52:42.6 & 4:34:36.0 & -24.08 &    2.9 & 0.15 &   46.5 &     74.0 &    0.062 \\
  16D33 &   1.00 &   0.76 &   0.95 &   0.7 &    1.5 & 285:45:47.8 & 4:37:39.7 & -22.08 &   11.3 & 0.31 &  117.3 &    163.6 &    0.096 \\
  16D34 &   0.34 &   0.40 &   0.66 &   0.3 &    1.2 & 285:52:56.3 & 4:39:50.1 & -22.75 &    1.1 & 0.11 &   22.5 &     45.0 &    0.051 \\
  16D35 &   0.60 &   0.42 &   0.66 &   0.4 &    0.9 & 285:56:16.0 & 4:45:04.3 & -24.08 &    1.8 & 0.14 &   29.4 &     58.9 &    0.060 \\
  16C1$\star$ & 0.61 & 1.44 & 1.06 &   0.9 &    8.2 & 285:52:01.3 & 4:23:24.4 & -25.40 &   87.7 & 0.51 &  196.5 &    246.7 &    0.446 \\
  16B1  &   0.53 &   0.63 &   1.02 &   0.5 &    2.8 & 285:51:30.2 & 4:42:12.5 & -23.41 &   11.8 & 0.28 &  108.1 &    140.7 &    0.109 \\
  16B2  &   0.44 &   0.56 &   1.01 &   0.5 &    2.6 & 285:51:58.4 & 4:46:00.9 & -23.41 &    8.4 & 0.23 &   89.7 &    118.3 &    0.093 \\
  16B3  &   0.77 &   1.25 &   0.94 &   0.7 &    1.3 & 285:47:12.5 & 4:44:34.4 & -23.41 &   10.3 & 0.32 &  123.0 &    174.5 &    0.084 \\
\end{tabular}
  \caption{$^{13}$CO(J=2--1) clump properties -- lower region. The meaning
  of the columns and the means by which the values were calculated are explained in
  \S\ref{sec:clump_properties}. Clumps marked by the asterisk at their label are
  excluded from the CMF, because they lie at the border of the observed area.}
  \label{table:13CO_clumps_lower_full}
\end{table*}


\begin{table*}
  \centering
\begin{tabular}{lrrrrrrrrrrrrr}
  Label &
  $\frac{\mathrm{FWHM}_{x}}{\mathrm{pc}}$ &
  $\frac{\mathrm{FWHM}_{y}}{\mathrm{pc}}$ &
  $\frac{\mathrm{FWHM}_{v}}{\mathrm{km\;s^{-1}}}$ &
  $\frac{R}{\mathrm{pc}}$ &
  $\frac{T_{\mathrm{peak}}}{\mathrm{K}}$ &
  $\frac{l_\mathrm{peak}}{\mathrm{deg}}$ &
  $\frac{b_\mathrm{peak}}{\mathrm{deg}}$ &
  $\frac{v_\mathrm{peak}}{\mathrm{km\;s^{-1}}}$ &
  $\frac{m}{\mathrm{M}_\odot}$ &
  $\frac{\Delta m}{\mathrm{M}_\odot}$ &
  $\frac{m_{\mathrm{vir}}}{\mathrm{M}_\odot}$ &
  $\frac{\Delta m_{\mathrm{vir}}}{\mathrm{M}_\odot}$ &
  $\frac{m}{m_{\mathrm{vir}}}$
  \\
  \hline\hline
  16E1 &   0.94 &   1.66 &   1.20 &   1.2 &    4.3 & 286:00:20.5 & 5:14:31.2 & -24.74 &  121.0 & 0.69 &  335.8 &    372.1 &    0.360 \\
  16E2 &   1.32 &   0.99 &   0.88 &   1.0 &    4.2 & 285:59:51.5 & 5:13:20.1 & -24.08 &   57.5 & 0.47 &  153.3 &    231.1 &    0.375 \\
  16E3 &   1.32 &   1.29 &   1.13 &   1.1 &    3.7 & 285:59:10.6 & 5:19:45.3 & -24.08 &   56.4 & 0.54 &  270.0 &    317.8 &    0.209 \\
  16E4 &   1.18 &   1.32 &   0.93 &   1.1 &    3.3 & 285:58:13.8 & 5:21:11.2 & -23.41 &   63.3 & 0.53 &  183.1 &    262.2 &    0.346 \\
  16E5 &   0.65 &   0.59 &   0.94 &   0.5 &    3.3 & 285:59:25.1 & 5:20:13.7 & -23.41 &   12.9 & 0.24 &   92.2 &    129.9 &    0.140 \\
  16E6$\star$&  0.32 & 0.48 & 0.75 &  0.3 &    2.3 & 285:56:44.8 & 5:11:26.8 & -23.41 &    3.0 & 0.13 &   34.5 &     60.9 &    0.087 \\
  16E7 &   0.52 &   1.10 &   1.02 &   0.7 &    2.0 & 286:02:28.5 & 5:11:39.4 & -24.08 &   15.8 & 0.34 &  132.9 &    173.2 &    0.119 \\
  16E8 &   0.34 &   0.97 &   1.00 &   0.5 &    1.8 & 286:02:57.9 & 5:13:47.5 & -23.41 &    7.0 & 0.23 &   90.9 &    120.4 &    0.076 \\
  16E9 &   0.64 &   0.34 &   0.75 &   0.4 &    1.5 & 286:03:41.9 & 5:16:38.3 & -23.41 &    3.0 & 0.15 &   40.1 &     71.5 &    0.074 \\
  16F1 &   0.61 &   1.06 &   1.03 &   0.8 &    3.2 & 286:01:01.6 & 5:08:48.7 & -23.41 &   23.8 & 0.38 &  153.5 &    197.6 &    0.155 \\
  16F2 &   0.81 &   1.63 &   1.04 &   0.9 &    2.8 & 286:02:13.3 & 5:09:02.6 & -24.08 &   31.1 & 0.44 &  181.6 &    232.0 &    0.171 \\
  16F3$\star$& 0.46 & 0.42 & 0.85 &   0.4 &    1.9 & 286:00:17.5 & 5:05:00.6 & -24.08 &    2.8 & 0.15 &   48.3 &     75.4 &    0.057 \\
\end{tabular}
  \caption{$^{13}$CO(J=2--1) clump properties -- upper region. The meaning
  of the columns and the means by which the values were calculated are explained in
  \S\ref{sec:clump_properties}. Clumps marked by the asterisk at their label are
  excluded from the CMF, because they lie at the border of the observed area.}
  \label{table:13CO_clumps_upper_full}
\end{table*}

\end{appendix}

\end{document}